# Light Pollution in USA and Europe: The Good, the Bad and the Ugly

**Authors**
F. Falchi[1]*, R. Furgoni[1], T.A. Gallaway[2], N.A. Rybnikova[3], B. A. Portnov[4], K. Baugh[5], P. Cinzano[1], C.D. Elvidge[6],

**Abstract**
Light pollution is a worldwide problem that has a range of adverse effects on human health and natural ecosystems. Using data from the New World Atlas of Artificial Night Sky Brightness, VIIRS-recorded radiance and Gross Domestic Product (GDP) data, we compared light pollution levels, and the light flux to the population size and GDP at the State and County levels in the USA and at Regional (NUTS2) and Province (NUTS3) levels in Europe. We found 6,800-fold differences between the most and least polluted regions in Europe, 120-fold differences in their light flux per capita, and 267-fold differences in flux per GDP unit. Yet, we found even greater differences between US counties: 200,000-fold differences in sky pollution, 16,000-fold differences in light flux per capita, and 40,000-fold differences in light flux per GDP unit. These findings may inform policy-makers, helping to reduce energy waste and adverse environmental, cultural and health consequences associated with light pollution.

**MAIN TEXT**

## 1 Introduction

Light pollution (LP), resulting from the alteration of natural night light levels by artificial light sources is one of the most evident pollutant in the Anthropocene ([1]), is continuously increasing in magnitude ([2],[3],[4]), notwithstanding, or, perhaps, due to the raising efficiency in producing light[5]. LP is a major environmental and health problem, known to be associated with depression, insomnia and other health disorders in humans ([6],[7],[8]) and potential changes in foraging, navigating and reproductive behaviour in wildlife species[9]. The widespread introduction of high intensity white light-emitting diodes (LEDs), praised by many for their high efficiency, does, in fact, only exacerbates the problem due to light emissions with "bluer" and more polluting light spectra compared to more yellow light emitted by previous lighting technologies, such as incandescent and low pressure sodium lights ([10],[11]). As a result, more short wavelength (commonly called blue) light, is introduced into the night environment.

[1] ISTIL - Istituto di Scienza e Tecnologia dell'Inquinamento Luminoso, Light Pollution Science and Technology Institute, Thiene, Italy
[2] Economics Department, Missouri State University, USA
[3] Remote Sensing Laboratory, the Center for Spatial Analysis Research, Department of Geography and Environmental Studies, University of Haifa
[4] Department of Natural Resources & Environmental Management, Faculty of Management, University of Haifa, Israel
[5] Cooperative Institute for Research in the Environmental Sciences, University of Colorado, Boulder, USA
[6] Earth Observation Group, Payne Institute, Colorado School of Mines, Golden, Colorado, USA

The technical parameters of light sources and actions required to lower artificial light at night (ALAN) pollution are well known ([12]), and some of them are already implemented in regional and national laws in several countries, including Italy, Slovenia, Chile, Spain, France and Croatia. These actions include: aiming the lights only downwards, instead of wasting light by directing it above the horizontal plane; orienting street lights towards the target (e.g. on the road or pathway, not towards private properties or windows), and turning lights on at the correct timing, using smart and adaptive lighting technologies. Other regulatory measures to reduce LP include regulating, on sound scientific basis, the absolute minimum lighting levels necessary to perform the action (e.g., driving or walking on a sidewalk) and using light sources emitting less impacting, blue poor spectra, while avoiding high intensity blue emission sources, such as e.g., white LEDs. The use of these strategies, suggested by light pollution experts, can reduce, by an order of magnitude or more, LP in heavily polluted areas.

This work presents the amount of LP produced by different geographic units, such as States and counties in the USA and NUTS2 and NUTS3 (*Nomenclature des Unités Territoriales Statistiques*, the French for Nomenclature of Territorial Units for Statistics) regions of the EU. It lists all the administrative units in Europe and USA from the best to the worst examples (in light pollution of the sky, in light emissions per capita and light emissions per income). This 'catalogue' will be useful to the scientific community and the policy-makers as a basis to explore the multiple causes of the more and less virtuous administrative units, helping to find better solutions to the global problem of light pollution.

For the analysis we use data on artificial night sky brightness from the New World Atlas of Artificial Night Sky Brightness ([13]), data on the flux emitted by light sources obtained from the VIIRS satellite images, the population densities and per capita income data for the above regional subdivisions, obtained from the Eurostat and the US Census Bureau. We use these data to calculate the following five measures for each administrative subdivision:

a) the percent of the area of an administrative subdivision with a given level of artificial night sky brightness (subdivided into 6 classes of ALAN pollution);
b) the percent of population living under a given artificial night sky brightness (subdivided into 6 classes of ALAN pollution);
c) the average artificial night sky brightness of the considered territory;
d) the artificial light flux per capita (FpC);
e) the artificial light flux per GDP unit, measured in US$ (FpD).

The FpC and FpD allowed to analyse LP in a new perspective, showing often that the most polluted areas, such as the metropolis are those that pollute less per inhabitants or per unit of income. These data also allowed to compare the different polluting power per capita and per income of the various cities and administrative units, finding that a similar light pollution can derive from cities with very different number of inhabitants.It is worth mentioning that the flux data is different from the radiance of the night sky given by the New World Atlas of Artificial Night Sky Brightness. The main difference for the purposes of this study is that the flux gives how much light is produced (and escapes to space) in each pixel area, while the second gives the artificial night sky brightness looking at the zenith from the centre of each pixel. The two things are related, but not in a trivial way. As an example, the flux coming from the Upper Bay in New York is essentially zero (no light sources are on the water), while the night sky observed from the centre of the bay is extremely light polluted, due to the lights coming from the surrounding sources. In fact, the World Atlas radiance data for each pixel was computed taking into account the lights coming from a circle of 200 km radius.

Combined with the abovementioned LP-reduction strategies, this knowledge can help to developed targeted policies aimed to achieve a substantial decrease in LP, thus *de facto* solving the problem and arriving to a sustainable use of artificial light at night. If we are unable to solve this problem, for which the countermeasures are well known, easy to implement, and can bring energy and direct and indirect monetary savings, along with benefits to biodiversity, health and culture, then our ability of solving more complex environmental problems, such as e.g., global warming, will remain in doubt.

## 2 Results

The analysis reveals extremely high discrepancies in LP (intended as artificial night sky brightness), in FpC and in FpD between areas in both the USA and Europe, with discrepancies, both absolute and relative, varying by magnitudes. On a smaller scale, the FpC and FpD was already investigated in 1997 for Italy[14].

### *2.1 Artificial night sky brightness*

Area and population statistics of the average artificial night sky brightness (in μcd/m$^2$) are reported in Tables S1-S4, separately for the State (Table S1) and county levels of the USA (Table S2) and for NUTS2 and NUTS3 of the EU (Tables S3-S4). In addition, distances from major metropolis to the nearest areas with the pristine skies are reported in Tables S5 and S6, which list the distance from the cities of over one million inhabitants to the nearest dark sky place (excluding dark skies in the sea), separately for the USA and EU. We computed the average artificial night sky brightness and assigned rank values, from darkest sky to brightest sky, to counties and to NUTS3.

Concurrently, in Figure 1 and Figure 2, we show the average artificial night sky brightness of NUTS3 EU regions and US counties. To enable cross-referencing, counties and NUTS3 in Tables S2 and S4 have the same colours as in Figure 1 and Figure 2. Lastly, Figure 3 and Figure 4 show the remaining pristine dark sky places in Europe and USA.

In Europe, Delft en Westland, in the Netherlands, has the highest night sky light pollution, with nigttime artificial sky brightness being almost 7000-times higher than that in the darkest NUTS3 region, Eilean Siar (Western Isles), in the United Kingdom. In the USA, the District of Columbia has the highest artificial night sky artificial brightness, being more than 200 thousand times higher than the darkest county, Yakutat City and Borough, Alaska.

### *2.2 Light flux per capita*

Figure 5 shows the flux per capita (FpC) in EU NUTS2 regions and USA States with the same colour key, in order to better compare Europe and USA. No table is given for flux per capita data at these administrative levels.
Concurrently, Tables S7 and S8 show the light flux per capita for EU NUTS3 provinces and US Counties.
Flux per capita at NUTS3 and counties level are visualized (with two different colour keys due to the higher average flux per capita in USA compared to Europe, so no direct comparison can be made using these two maps) in Figure 6 and Figure 7.

As Figure 8, which features FpC histograms for European Union 28 countries, shows, Germany has an average of about half of FpC than the rest of Europe and only two of its regions --Altötting and Kaiserslautern – are outliers with about three times more flux per capita than Germany's average. The United Kingdom has the second best average level of FpC in Europe, while France has a national average slightly higher than the rest of Europe, and Spain and Italy have much higher FpC than the rest of the EU countries.
Figure 9 shows the histograms of the number of counties in USA and 5 representative States for each FpC range, along with the national average that is almost three times higer than Europe's and over five times higer than Germany's.

In Europe, Pohjanmaa, in Finland, has the highest flux per capita as detected by the satellite, while Schaffhausen, Switzerland, has the lowest. The differences in the light flux between these regions are so big that 120 Schaffhausen's inhabitants are needed to produce the same flux detected by satellite for a single Pohjanmaa inhabitant.
In the USA, Loving County, Texas, scores 16000 times more flux per capita than New York County. Part of this discrepancy can be explained by the oil wells present in the area of Loving County. Kalawao County, Hawaii, does not show on the satellite data, so we have no detected flux for this county. Curiously, the least (Kalawao) and second to least (Loving) counties in the USA are at the opposite extremes of the FpC ranking.

### *2.3 Light flux per GDP unit*

Table S7 and Table S8 report ranking of the USA States and Counties, and European NUTS2 and NUTS3 regions in terms of the flux per US$1 (FpD) of the average per capita income for each county and region. The counties and NUTS3 regions in these tables are ordered from the lowest flux per dollar to the highest, as appears in the 'Rank FpD' column.

As Tables S7-S8 show, the regions ranked *best*, are generally densely populated non-rural areas. The population sizes of these regions may be below average but their population densities are usually above average, and they consistently have mean income levels that are well-above average, often by a factor of 2 or 3. The regions ranked *worst* in terms of FpD are more varied. Such regions are usually non-urban. Incomes and populations in these regions can be above or below average, although population density is nearly always well below average.

Because people and economic activity go together, there is a correlation between these rankings and the rankings based on flux per capita. In general, small populations require more infrastructures and carry large distances between homes and places of employment, which can help to explain a high FpD, while large populations are not as important in explaining low FpD. For these best performing regions, high levels of income are a more important explanation for relatively high light emissions. In the regions performing worst, income may vary from below average to above average, with a little impact on the rankings. For example, a high flux from the much of the central and western USA (i.e., the region that is comparatively dark), has more to do with small population than with low incomes.

The FpD ratios for Europe show that only 2 of the top 10 regions and 6 of the top 20 regions are in the top 10 and top 20 of the lowest light flux per capita rankings. For the USA, the two rankings are more consistent but still only 7 of the top 10 and 13 of the top 20 are found among the regions ranked best in terms of flux per capita. By contrast, the rankings of the worst performers are more correlated. Thus, In Europe, we find 9 of the 10 worst and 16 of the 20 worst regions for FpD overlap with the worst rankings for FpC. For the USA, these numbers are 9 of the bottom 10 and 19 of the bottom 20. The worst four regions for Europe and the worst three for the USA are identical when comparing FpD and FpC. That is, as an alternative way of sorting, FpD gives distinctive rankings for the best regions, but not for the worst regions.

### *2.4 Global ranking*

We ordered the smallest administrative units in the USA and EU, using the above mentioned three parameters: the average (over the unit territory) artificial night sky brightness at zenith, the flux per capita (FpC) and the flux per GDP unit (FpD). Each of these criteria gave us a rank from the best to the worst performer in terms of LP. It is worth mentioning that being at the top of one or two rankings does not imply to be at the top of the other(s). An extreme case is the New York county, which is second best in the nation in terms of FpC and FpD ranks, i.e. it has the second lowest FpC and FpD in the USA, but it ranks 3141th (that is second to last) in terms of the mean artificial brightness. In other words, this county has the second worst sky in the nation. Similar is the case with Paris in Europe, which ranks 5$^{th}$ in terms of FpC, 1$^{st}$ in FpD, but is ranked 1363th in the average sky brightness ranking, i.e. is 3$^{rd}$ among regions with the worst overall sky brightness. This can easily be explained by extremely high population densities in the big cities that make them appear virtuous in terms of FpC and FpD, but overall wasteful in LP due to extremely large numbers of people living there. The really virtuous are those counties and provinces which have low population densities and a low emitted flux. A fair comparison should be made by choosing cities with similar population sizes. In this respect, comparing Munich in Germany to Turin and Milan in Italy, we found that the Italian cities has three to four times more FpC than German cities of a comparable size. By the same token, Rome and Madrid have four times FpC than Berlin and Paris.

The global rank is reported in Tables S7 (EU NUTS3) and S8 (US Counties) for all US counties and European NUTS3. The Tables 1 and 2 report, ordered by this global rank, the top 25 (the good), the last 26 to 50 (the bad) and the worst 25 (the ugly) between EU NUTS3 and US Counties. Between the most virtuous the top 3 on the list in the USA are Hoonah-Angoon Census Area, in Alaska, Catron County, in New Mexico, and Jeff Davis County, in Texas, while in Europe are Ammerland, in Germany, Bornholm in Denmark and Northeim in Germany. The worst three in USA in this global ranking are McKenzie County in North Dakota, McMullen and Karnes Counties in Texas, while in Europe are Delft en Westland and Oost-Zuid-Holland in the Nederlands and Pohjanmaa, Finland. Figure 12 shows in green the TOP 25 'good' NUTS3 (Upper panel, Europe) and TOP 25 counties (lower panel, USA) in this global rank, and the worst 50 in red (the 'ugly' last 25 places) and orange (the 'bad', 26th to 50th place from bottom).

## 3 Discussion

### *3.1 Artificial night sky brightness*

The three provinces with the best sky in Europe are all in Scotland. Of the best 25 NUTS3 (that is regions with the darkest skies), 5 belong to Austria, 4 to the UK, 3 are in Latvia, 2 are in Lithuania and Bulgaria each, 1 in Denmark, Estonia, Greece, Iceland, Norway, Romania, Spain and Sweden. In USA, the three darkest skies counties are in Alaska (which hosts 12 counties in the first 25 places). Other US states with the least light polluted counties in the top 25 are: Montana (3), Oregon (3), New Mexico (2), Utah (2), Nebraska, Nevada and Idaho (1 each). It is to be noted that, on the overall, the USA has darker skies than Europe, and, in fact, the best European NUTS3 would be only in the 120th place between the US Counties in terms of mean radiance.

On the negative side of the nighttime sky brighness scale in Europe, we found two provinces in the Netherlands. In particular, the brightest night sky is in Delft on Westland, where probably most of the light is produced in greenhouses to speed up plant growth. This gives to the province more than twice the brightness of the second, L'Aia metropolitan area, and more than three times that of the third most light polluted area in Europe, Paris. The worst 25 light polluted places are found in the UK (10), the Netherlands (5), France (4), Italy (2), and one each for Belgium, Poland, Romania and Spain.

In the USA, the most polluted skies are in the District of Columbia, New York county and Hudson county. Among the worst 25 LP places, we found 6 counties in Virginia, 5 in the New York state, 3 in New Jersey, 2 in Illinois and Texas, one in each of District of Columbia, Maryland, Massachusetts, Minnesota, Missouri, Pennsylvania and Wisconsin states. The ratio in the sky pollution between the most polluted and least polluted counties is over 200,000 in the USA, and about 7,000 in Europe for the NUTS3. This difference is mostly due to the fact that the darkest US counties are much less polluted than the darkest European regions, while major metropolitan centers generated about the same amount of artificial night sky brightness on both sides of the Atlantic.

### *3.2 Light flux per capita*

As Figure 5 shows, it appears that on the overall, the USA produces more light per capita than Europe. This extends the result found comparing USA to Germany ([15]). This can be explained, even in presence of national norms (e.g. IESNA RP-8) that generally prescribe lower lighting levels to light up roads, compared to the EU norms (e.g. EN 13201), with the fact that US roads are on average much wider than European and serve more single-family homes in the suburbs where population density is much lower compared to that in Europe. Moreover, the share of light consumed by private users in the USA is also higher than in Europe[16]. Therefore, Figure 6 and Figure 7 use different color scales and cannot be directly compared.

The regions of Pohjanmaa in Finland, Finnmark and Troms in Norway are three most proliferative provinces in Europe in terms of FpC. Other countries in Northern Europe also lead in terms of light flux, with 11 regions in Finland, 5 in Norway and Sweden each, 2 in the Netherlands, and 1 in Iceland and UK 1 each forming the list of 25 most polluting NUTS3 in Europe. The three more virtuous European regions in terms of FpC are Schaffhausen in Switzerland, Oldenburg in Germany and Schwyz in Switzerland. Over one hundred (120) inhabitants of Schaffhausen are needed to produce the same light flux of a single resident of Pohjanmaa. Of the top 25 most virtuous provinces, 21 are in Germany, 2 in Switzerland and 1 in the United Kingdom and France each.

Europe's flux per capita shows a geographic gradient of increasing LP per capita toward the southern countries of Portugal, Spain, southern France, Italy, Croatia and Greece and also towards the northernmost provinces of Finland, Iceland, Norway and Sweden. At least in part, this gradient in southern and northern directions can be explained by cultural differences and habits between countries. It is also possible that high flux per capita detected in northern countries may be explained with the fact that when the satellite passes, sides of roads may be covered by snow, producing a much higher flux escaping toward the outer space. An additional bias can also be due to the effect of aurorae that rises the light detected by satellite. This problem was taken under control by using a special GIS mask, that of the stable light as detected by DMSP satellite, that cut all the aurora's background where no permanent artificial lights are present. We checked the sky brightness values as detected by the available SQM (Sky Quality Meter) measurements[17], finding that, on average, the measured value is 0.1 magnitude per arcsecond squared darker than the values given by the World Atlas. This means that the flux we used for the computations can be off by about 10% at most. In fact, we also checked one of the highest flux per capita area in Sweden and found that there were brightly lit roads crossing the forests without any village in the surroundings. This points out that the dark skies that are left in the Scandinavian peninsula may be more due to low densities than to the virtuosity of the light pollution control. Therefore, there are substantial margins to reduce light pollution and to extend again the area of pristine skies.

Most virtuous in terms of LP are the countries in the central-east part of Europe and part of UK (England and Wales), from which LP gradient shows an upward trend towards the south and north. Interesting observations can also be made on singular nations. Although Germany has a relatively low FpC, large differences are evident between the former East and West Germany, with the former exhibiting higher FpC than the latter. The former Czechoslovakia is also split in two, with the Czech Republic producing more FpC than Slovakia. In Italy, the German speaking SudTirol-Alto-Adige province has a FpC similar to those in Germany and Austria, being much smaller than in the rest of Italy. A comparison with another mountain province in the Italian Alps, Valle d'Aosta, shows that this province has three times more FpC than the SudTirol-Alto-Adige.

### *3.3 Light Flux per GDP unit*

Another important factor in examining light pollution is the level of economic activity. In economic terms, FpD of income can be viewed as a cost-benefit ratio. As with other forms of pollution, such as e.g., carbon emissions, this is an important consideration. A region that generates twice as much LP but generates ten-times greater per-capita income as other regions can be arguably considered cleaner and more efficient.

When considering FpD maps (see Figure 10 and Figure 11), we see drastic changes compared to more familiar maps that show only the light flux. If the effects of population and economic activity on artificial sky brightness were constant, then flux to per-capita incomes would have a homogenizing effect, with urban and rural areas, as well as well-to-do and poor areas, tending to look more similar in terms of LP. Instead, the results point out at dramatic reversal in rankings. Indeed, after normalizing by regional incomes, urban areas seem more virtuous in LP and rural areas become to look much less virtuous. This reversal has a simple explanation. In economic terms, road lighting is a public good. For any given light fixture, multiple people can use the light without one person's consumption necessarily interfering with another. Similarly, there are a well-known agglomeration effects, internal and external economies of scale, and network externalities that explain the greater-than proportional level of economic activity in urban areas. Accordingly, the average fixture provides lighting to more people and facilitates more economic activity in areas with greater population density.

However, urban-rural densities and economies of agglomeration do not explain all the changes in ranking. There remain important differences that suggest that some areas do better than others even after controlling for economic activity and population. Determining how much of these differences is due to local regulations, historical path dependency, cultural preferences or better lighting design must be sorted out through future research. A few examples will serve to highlight these discrepancies which may justify future investigation.

While Paris's top ranking in the FpD hierarchy can be explained by its high level of affluence and very high population density, the same cannot be said about Schaffhausen, which ranks third among least light polluting places despite having only about 55% of Paris's mean income and less than 2% of Paris's population density. Similarly, the above mentioned Netherland's Delft en Westland is among the worst performers despite being a a densely populated urban area with above average per capita income.

Systemic differences between countries are even more important. Germany, for example, consistently does very well in all the rankings, while Portugal and the USA tend to perform poorly. Indeed, we can find large and consistent differences between, for example, Germany, the UK, France, Italy, and Spain. When compared to each other, they can be ranked from best to worst in that order. This ranking holds true whether we compare the average FpD for urban regions, rural regions, or *peri*-urban regions. The one exception is that France's urban regions do slightly better in terms of LP than the regions of the UK. These country-wide differences highlight the fact that demographic and income factors alone cannot explain all the variation between the "good" the "bad" and the "ugly."

Finally, it is worth noting that while looking at FpD or FpC can help highlight the actual *causes* of LP, such as population sprawl, overextended infrastructures and lenient national standards towards nighttime illumination. However, FpC is im fact the true measure that determines the actual damages to human health and wildlife. Indeed, the costs of such damages are multiplied, not divided, in areas with greater concentrations of people and economic activity.

### *3.4 Global ranking*

With regard to artificial night sky brightness, light flux per capita, and light flux per GDP, we have documented the good, the bad, and the ugly. These findings should prove useful to policy makers and scientists alike. By specifying both the magnitude and location of light pollution, these results could facilitate analysis of both the causal and mitigating factors that lead to higher or lower levles of light pollution in various administrative units. Likewise, the findings in this paper could help with the study and mitigation of problems commonly associated with light pollution including wasted energy and harmful impacts on aesthetics, human heath and the environment ([7],[18],[19]). For example, a variety of papers have examined various deleterious effects of LP by using satellite data, local observation, and controlled experiments ([7],[20],[21],[22],[23]). The economic factors contributing to light pollution have also been examined ([24],[25]). This paper's results, however, are uniquely useful in that they provide geographic, economic, and light pollution data and comparitive rankings for thousands of administrative units in both Europe and the United Sates.

These results should help open doors for continued research and analysis in areas such as examining the impact of LP on insects, birds, bats, sea turtles, rodents, and human health ([7],[23], [26],[27],[28],[29],[30],[31],[32]). Importantly, such research is tied not only to exterior lighting, but also includes such things as gas flaring ([33]). Similarly, as scientists learn more about the health effects and wasted energy associated with artificial lighting, better interior lighting is a growing concern ([34],[35]).

### *3.5 Limitations*

Among the limitations of the data produced by this study is the need to sort out the statistical magnitude and significance of various demographic and economic explanatory variables on the level of local light pollution. More detailed LP data, with respect to different spectra, would also be useful. All of these concerns are areas of ongoing and future research.

The light flux and light pollution generated by regional units, either per capita or GDP unit, may be a function of different factors, including climatic conditions, latitude, economic structure, population and production density, the level of infrastructure development, urbanization patterns and many others. A future analysis should thus attempt to investigate these multi-faceted links between light pollution and light flux, on the one hand, and an array of locational and economic factors, on the other, by applying multi-variate statistical analysis tools

It should be noted that our work is based on what the satellite detects, and the detected light can be from electric lighting or gas flares or even aurorae. Gas flares coming from oil and gas wells have, beside visible light emission, a strong near infrared flux, where VIIRS is very sensitive. For this reason, in areas where these emissions are present (e.g. McKenzie and surrounding counties in North Dakota), the light flux is overestimated. Even in the subset of electric lighting, we cannot distinguish public lighting from private, road lighting from sport fields, from industrial plant, from greenhouses and so on. Nor this is the aim of this work. We put at disposal of the scientific community and the politicians the data we computed. It is up to who want to use this data in specific places to investigate the causes of a good or a bad position in the rankings.
By the way, the fact that an high light flux per capita derives from a waste in public lighting or the presence of a big industrial plant does not change the fact that the flux per capita in that region is high. The problem for the night environment does not change.

## 4 Materials and Methods

For the present analysis we used several datasets, as detailed below:

The light flux (from radiance detected by Suomi NPP satellite), artificial night sky brightness at zenith (form the New World Atlas), population density (Landscan), per capita income (OECD's online regional database and States Census's American Community Survey), vector files of the European NUTS administrative subdivisions[36] and US states and counties[37].

The light flux was obtained from the dataset of the radiance detected in 2014 by the VIIRS on Soumi NPP satellite prepared as described in the New World Atlas paper, by taking into account the area of each 30"x30" latitude-longitude projection at different latitudes. The units used for calculation are somewhat arbitrary, simply obtained by multiplying the radiance of the VIIRS dataset (in nW $cm^{-2}$ $sr^{-1}$) by the pixel area measured in square kilometres, obtaining the dimensions of a radiant intensity in $10^{-7}$ W $sr^{-1}$. To obtain the correct radiant flux emitted above the horizon, assumptions should be made on the upward emission function (for a Lambertian emitter, simply multiply by $\pi$), on the average lamp spectra, while in order to get the light flux produced by light sources, additional assumptions should be made on the light intensities emitted at different angles by the sources, on the reflection by lighted surfaces and on the screening effect of obstacles.

The zenith artificial night sky brightness was taken from the supplement to the New World Atlas of Artificial Night Sky Brightness[38] that has the same resolution and projection of the light flux dataset. It is worth noting that the World Atlas is substantially different from the light flux dataset. In fact, the light flux is simply how much light is produced in a place (derived from the radiance detected from satellite), while the World Atlas depicts the consequences on the night sky brightness at the zenith in each site on Earth produced by all the lights emitted in a radius of about 200 km. A site with no sources (i.e. black in the light flux dataset) can be nonetheless heavily light polluted if its surroundings are full of sources. The zenith sky brightness of the World Atlas was computed using a light propagation model in a standard atmosphere, taking into account for the altitude of the sites and of the Earth curvature.

The 30"x30" latitude-longitude population data was taken from the LandScan™ 2013 High Resolution global Population Data Set copyrighted by UT-Battelle, LLC, operator of Oak Ridge National Laboratory[39].

The European economic data came from the Organization for Economic Cooperation and Development (OECD). Data were downloaded from the OECD's online regional database[40]. The data include the estimates of 2014 regional income per capita for each NUTS3 region, stated in 2010 US dollars and calculated using purchasing power parity. Similarly, the US data were obtained from the United States Census's American Community Survey (ACS). Data were downloaded from the US Census Bureau's American Fact Finder website[41]. Data was obtained for every county, or county equivalent, in the USA. These data also show 2014 per capita income in real US dollars, but use 2014 as the base year.
Both of these data sets were cleaned up in Excel and then imported into ArcInfo. Two join operations were done to combine these data with the light pollution data for each NUTS3 region in Europe and each county in the USA. The resulting maps and dataset were then used for further analysis, including sorting and ranking by different attributes as well as the creation of additional maps that show flux per capita and flux per dollar of income.

All the datasets were analyzed in the open access Geographic Information System QGIS Desktop 2.16.3 (https://qgis.org). The maps were produced in QGIS.

## 5 Conclusions

In the present analysis we found that there are great differences in the studied parameters between Europe and USA, with, e.g. USA haveing almost three times the Flux per Capita compared to Europe. We also found differences between countries inside EU (e.g. Portugal with four times the FpC of Germany) and USA (e.g. South Dakota with five times the FpC compared to New York). Greater differences are found between smaller administrative units, in part due to differences in population densities, presence of industrial plants, but also due to different lighting habits with Germans used to light less their cities compared to Portugal, Spain, Italy and Greece.

Also very interesting are the discrepancies in the Flux per Income showing that a direct correlation between the wealthy of a country or a state and the light it uses at night does not exist. There is evidence of the contrary inside Europe and inside USA, with central Europe far more virtuous (i.e. it pollutes less per unit of income) compared to southern countries. Also northern countries are less virtuous, but this may be due, in part, to a possible overestimation of the light flux produced (e.g. due to reflectance of snow or stray lights and northern lights possibly detected by satellite and not completely filtered out).

We hope that this work will be of great stimulus to politicians to lower the impact of their counties, states, provinces, regions, landers, countries by taking as examples to follow those best ranking in our score: the 'Goods'. Also, examining in depth the causes of the worst scores, the 'Bads' and the 'Uglies', will be of help in paving the way toward a more sustainable night lighting.

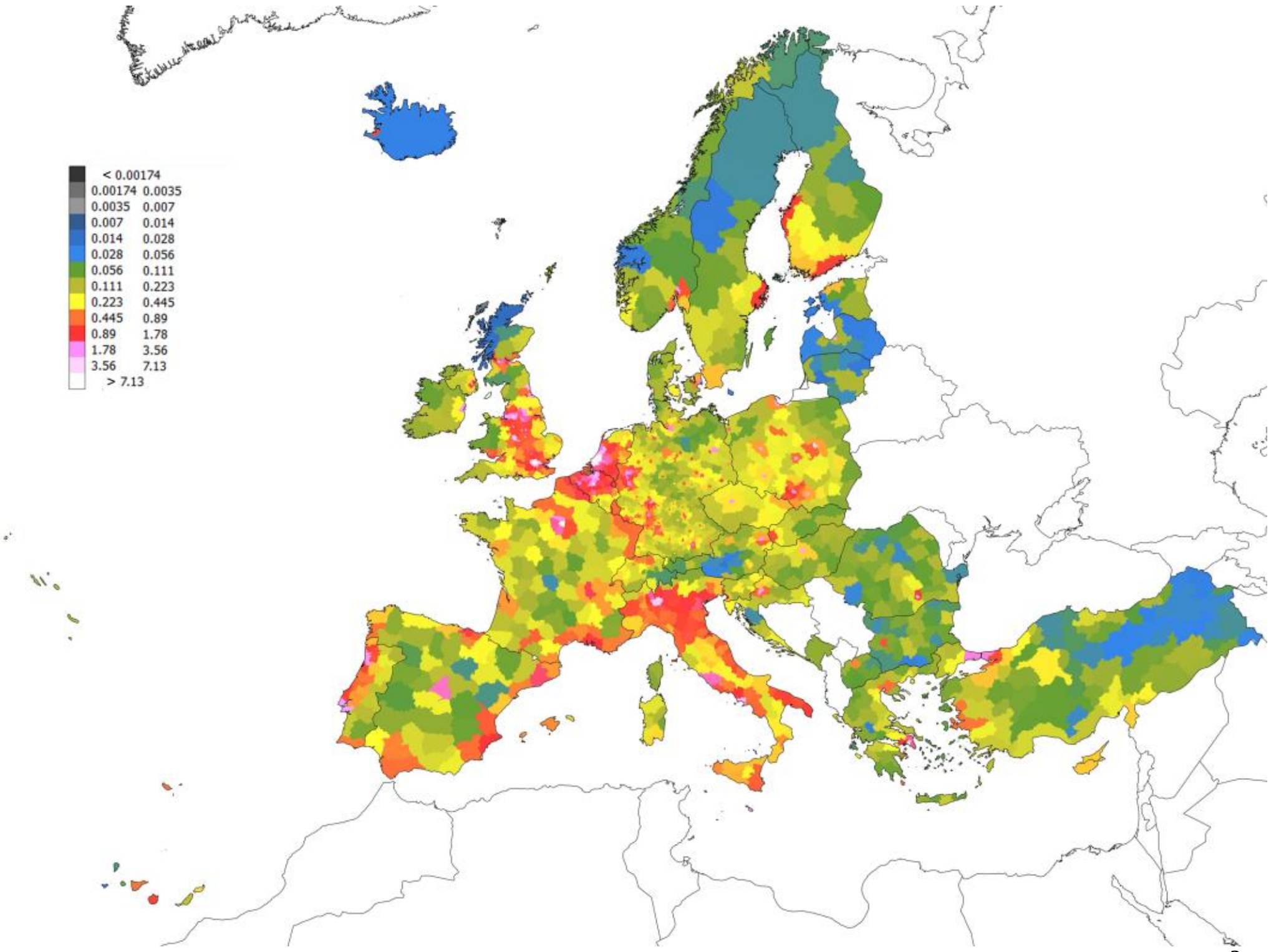

**Figure 1. Average zenith artificial night sky brightness of each NUTS3 region in Europe (mcd/m$^2$). The pollution of the clear night sky doubles at each change of colur, starting with pristine uncontaminated conditions (black colour, not present in Europe), to the white of the brightest metropolis. Note that these are data averaged over the surface of the provinces so that Madrid and Rome, for example, benefit of the relatively large surfaces of their provinces, compared to the small area of the Paris and London NUTS3 counterparts.**

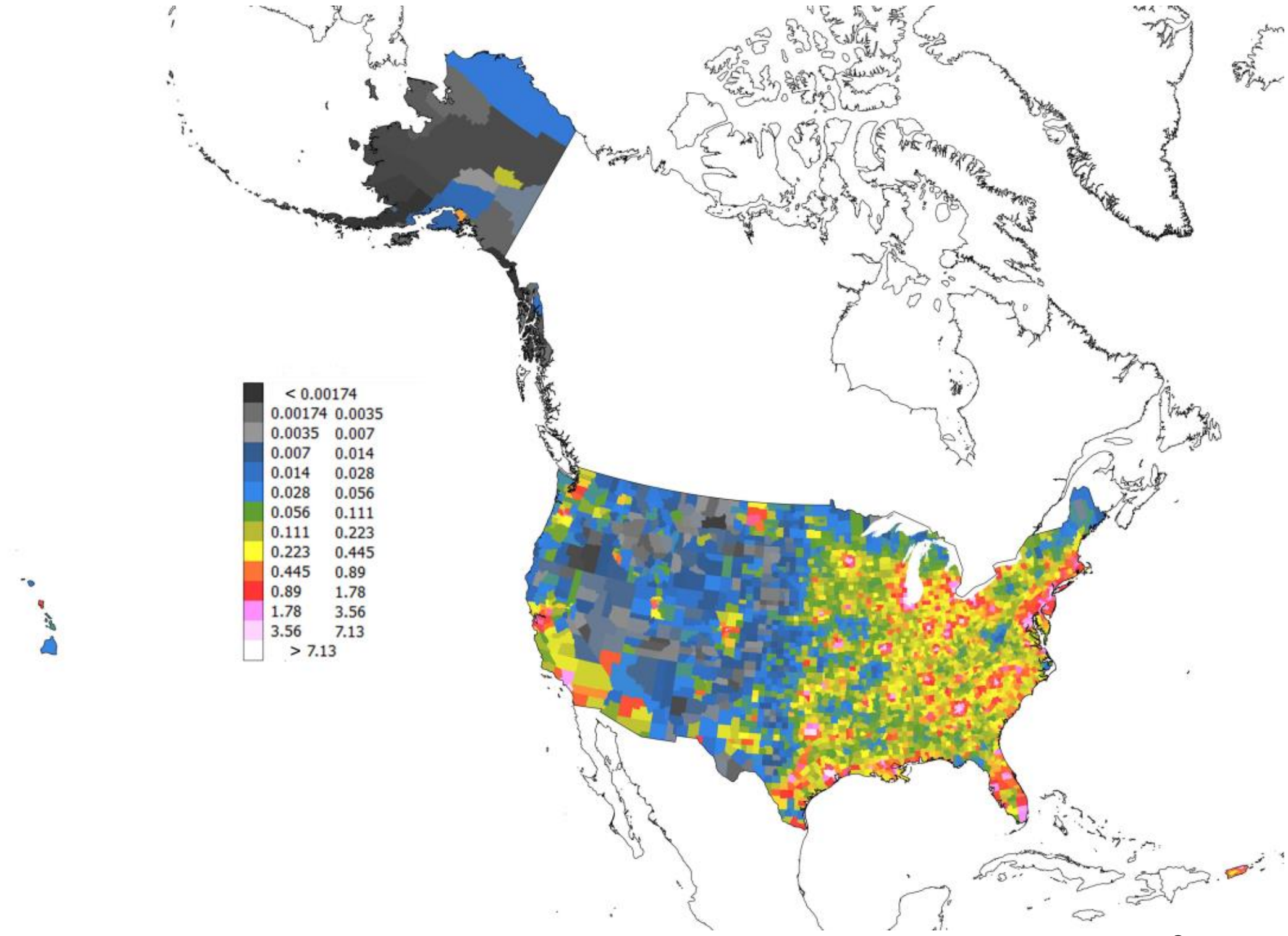

**Figure 2. Average zenith artificial night sky brightness of each county in USA (mcd/m$^2$). The pollution of the clear night sky doubles at each change of colur, from black to white.**

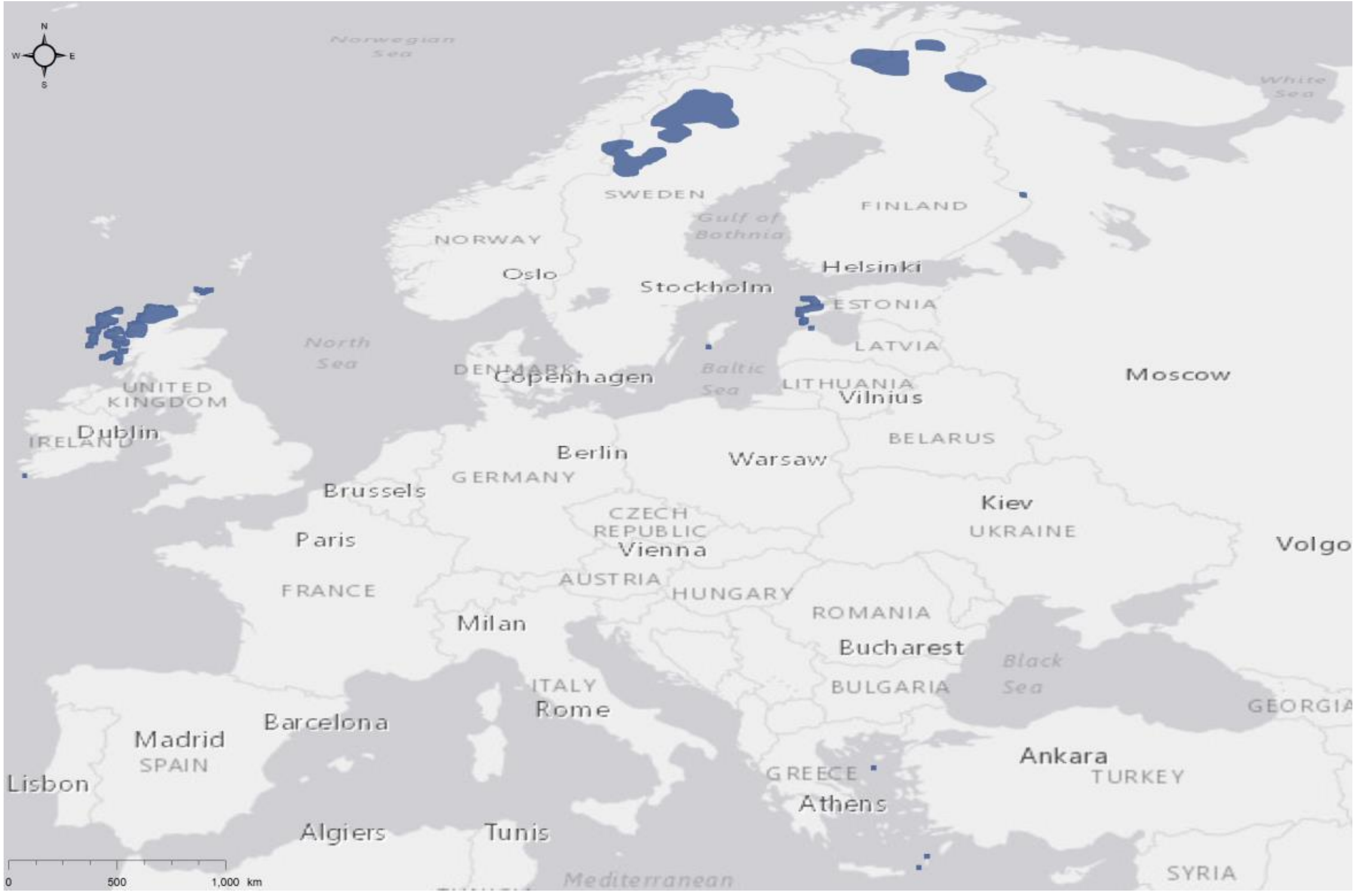

**Figure 3. Remaining areas with pristine skies in Europe (light blue colors mark spots with artificial brightness of up to 1% above the natural light, less than 1.7 μcd/m$^2$). Iceland presents a great part of its territory in pristine conditions (not shown here).**

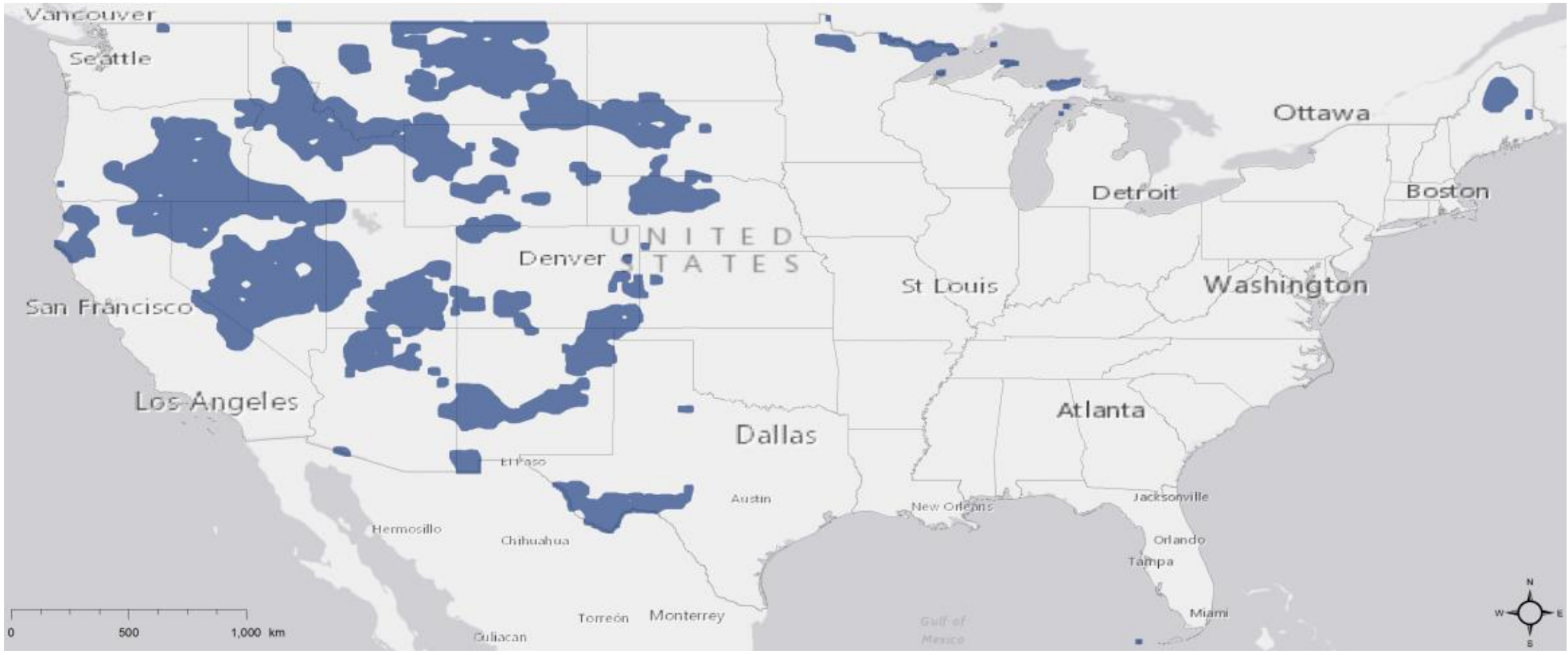

**Figure 4. The remaining areas in pristine sky in the continuous continental USA states (light blue colors mark spots with artificial brightness of up to 1% above the natural light, less than 1.7 μcd/m$^2$). Most of Alaska and some of the Hawaii islands have pristine sky conditions (not shown here). Note the differences with figure 2, where the main city inside each county rised the average of the sky brightness of the county itself.**

Note: Light Grey Canvas (Esri, HERE, DeLorme, MapmyIndia, © Open Street Map contributors, and the GIS user community) was used as the basemap

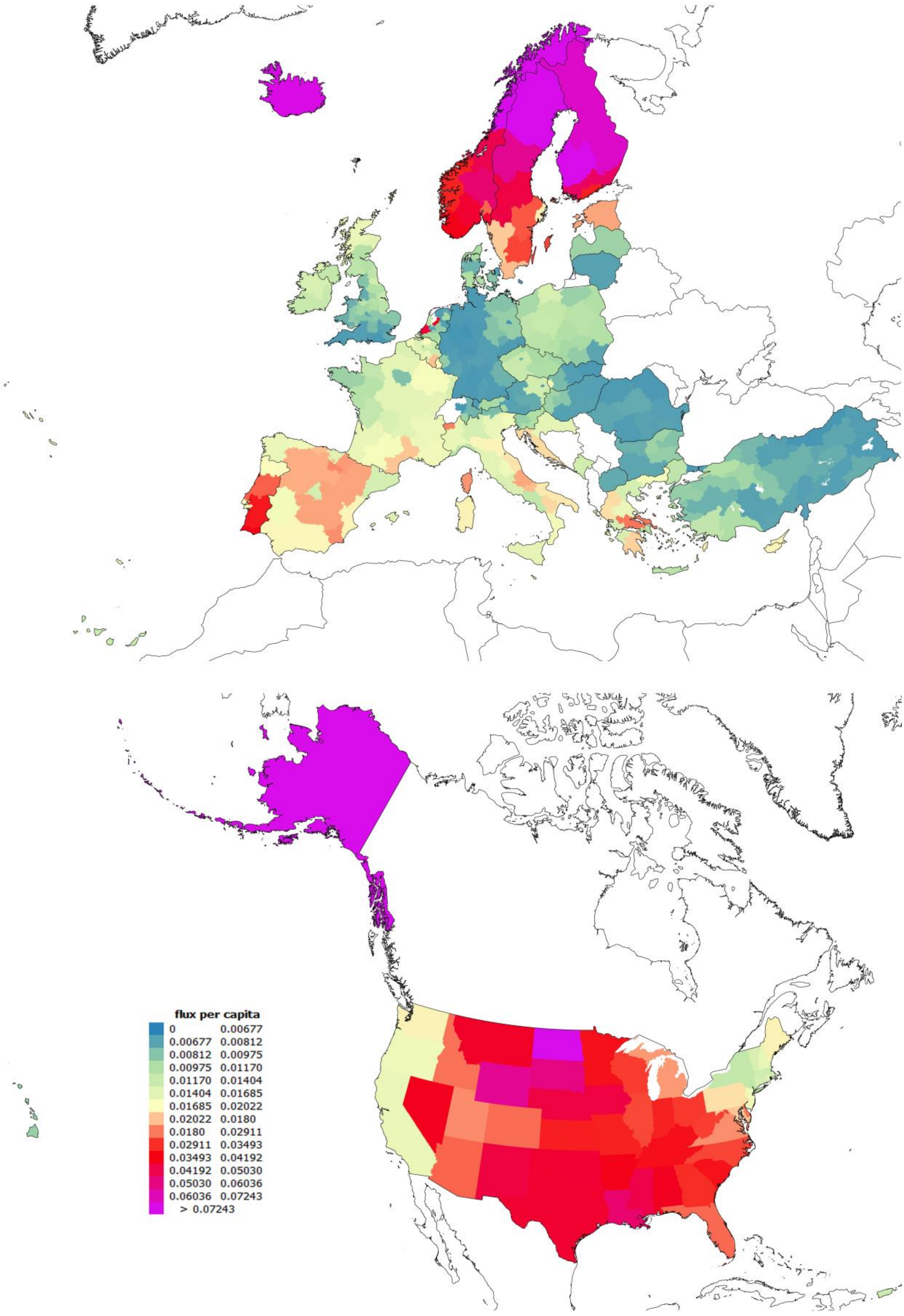

**Figure 5. Flux per capita levels at EU NUTS2, upper panel, and USA States, lower panel (in arbitrary units, ∝ W $sr^{-1}$ per capita, see main text). In these maps EU and USA are directly comparable, showing that most of the USA, except north-east and Pacific states, produces far more light per capita that most of Europe where the less polluting are mainly located in central Europe.**

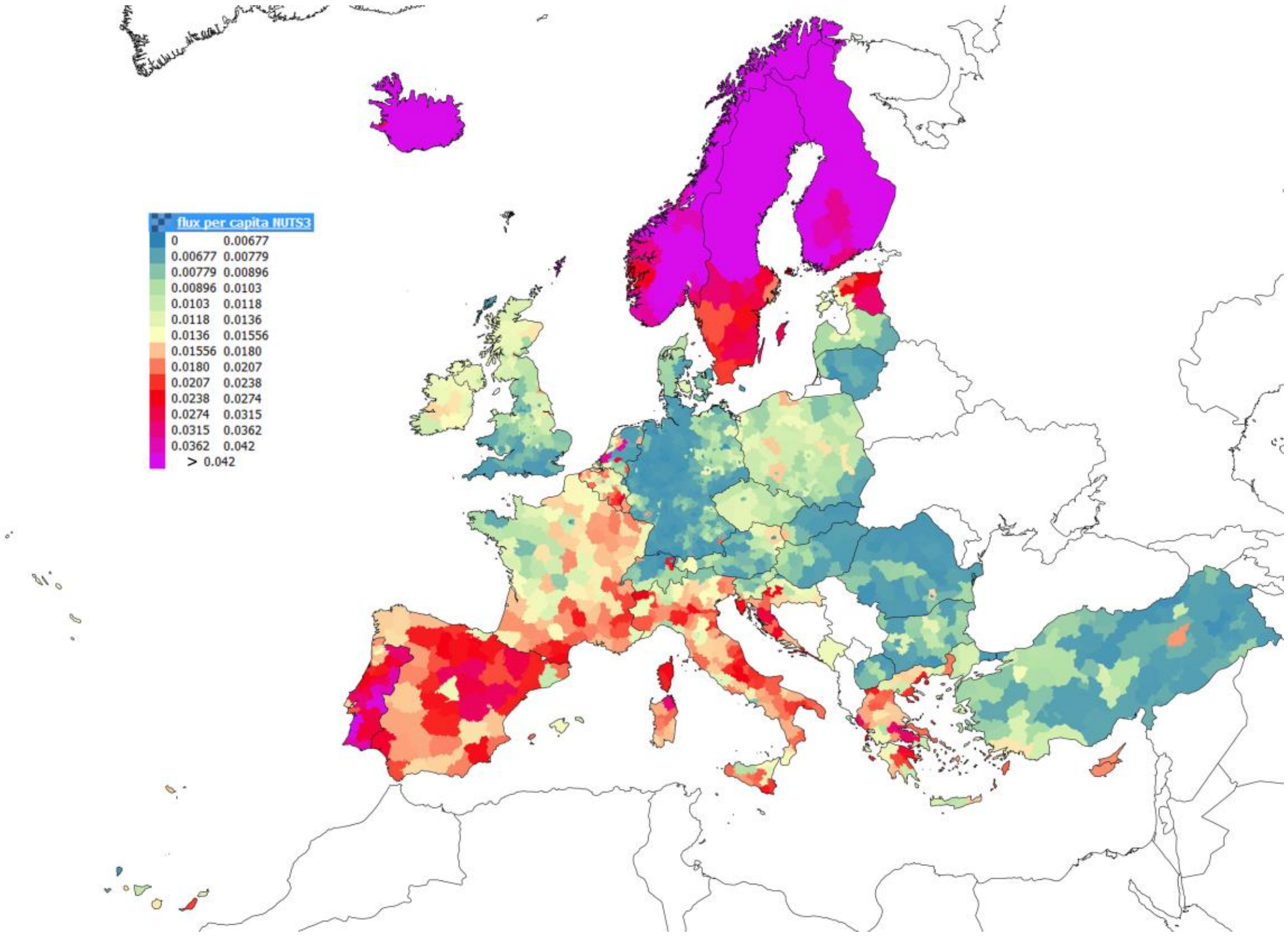

**Figure 6. Flux per capita (in arbitrary units, ∝ W sr$^{-1}$ per capita, see main text) at EU NUTS3 province level. Beware that the key is different from Figure 7. Each colour mean half the flux per capita compared to USA.**

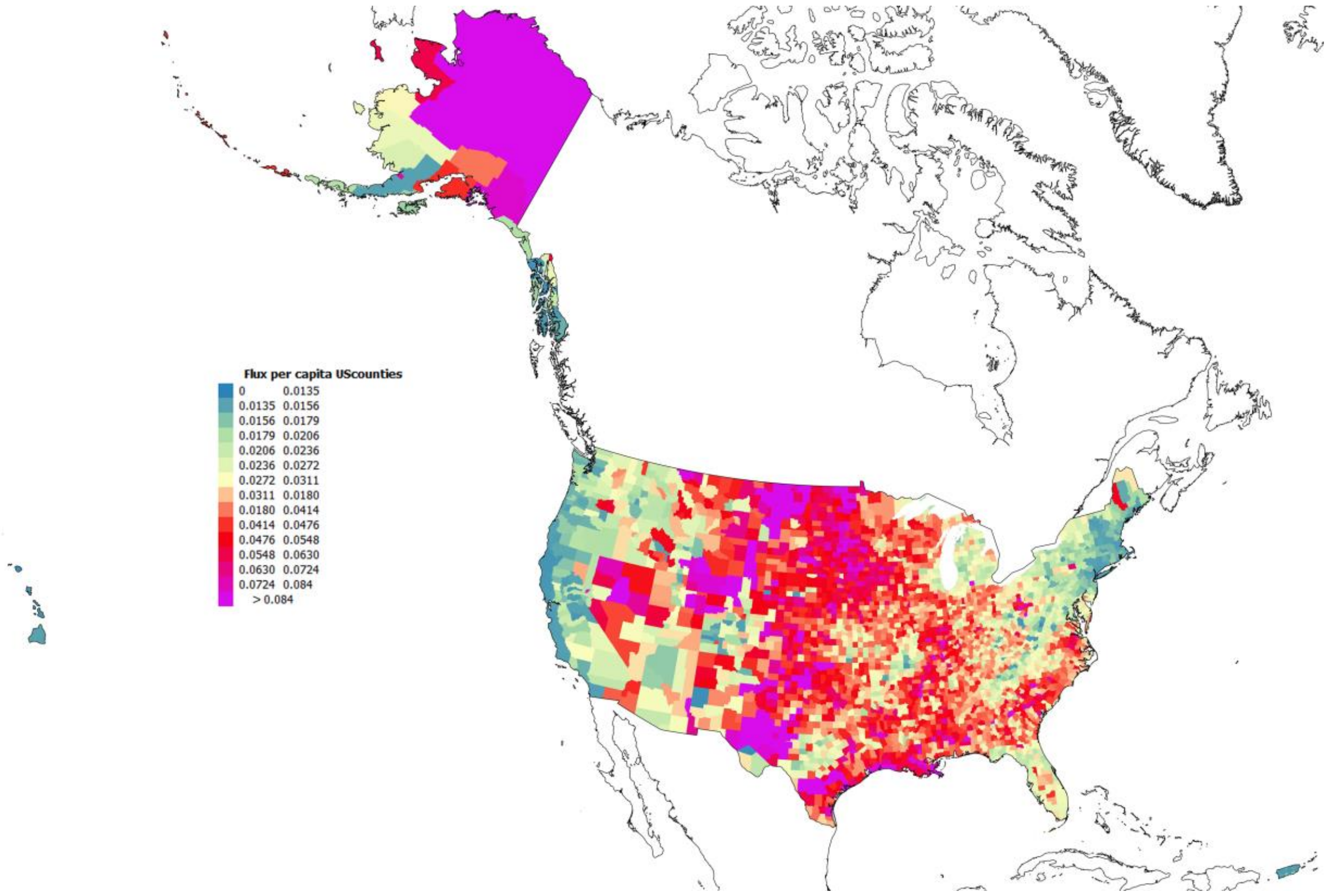

**Figure 7. Flux per capita (in arbitrary units, ∝ W sr$^{-1}$ per capita, see main text) at US Counties level. Beware that the key is different from Figure 6. Each colour mean twice the flux per capita compared to Europe. This allowed to show better the differences between counties in USA, otherwise most of them would have been compressed into the highet levels of pollution.**

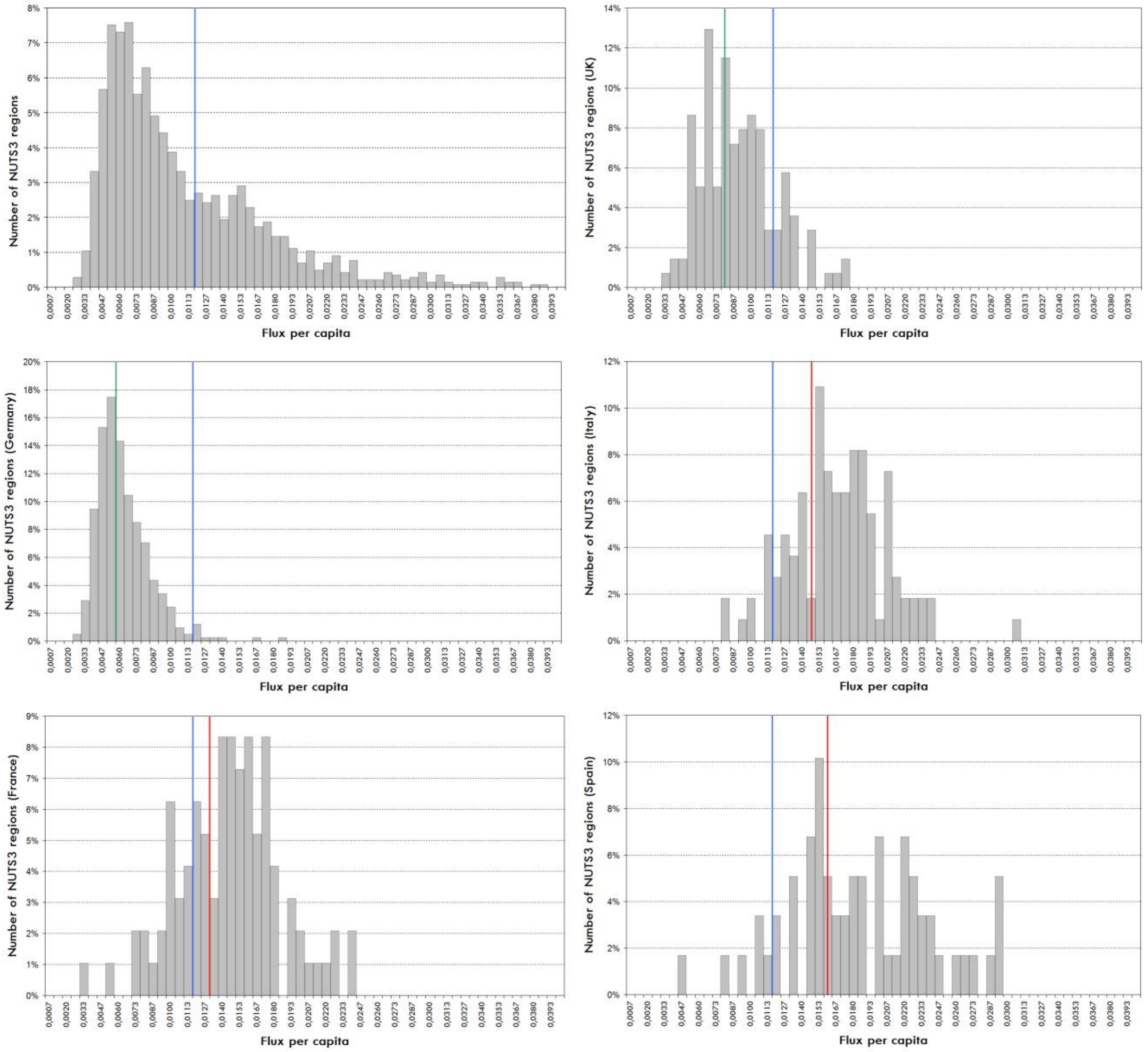

**Figure 8. Flux per capita (in arbitrary units, $\propto$ W $sr^{-1}$ per capita, see main text) in all the countries where NUTS are defined (including also non EU countries) (upper left graph), and in the main 5 European countries. The blue line indicates the average EU 28 flux per capita. The green (if better) or red (if worse) lines indicate the country's average. The histograms show the percent of NUTS3 in EU and United Kingdom, Germany, Italy France and Spain in each flux per capita range. From left to right the pollution produced per capita increases.**

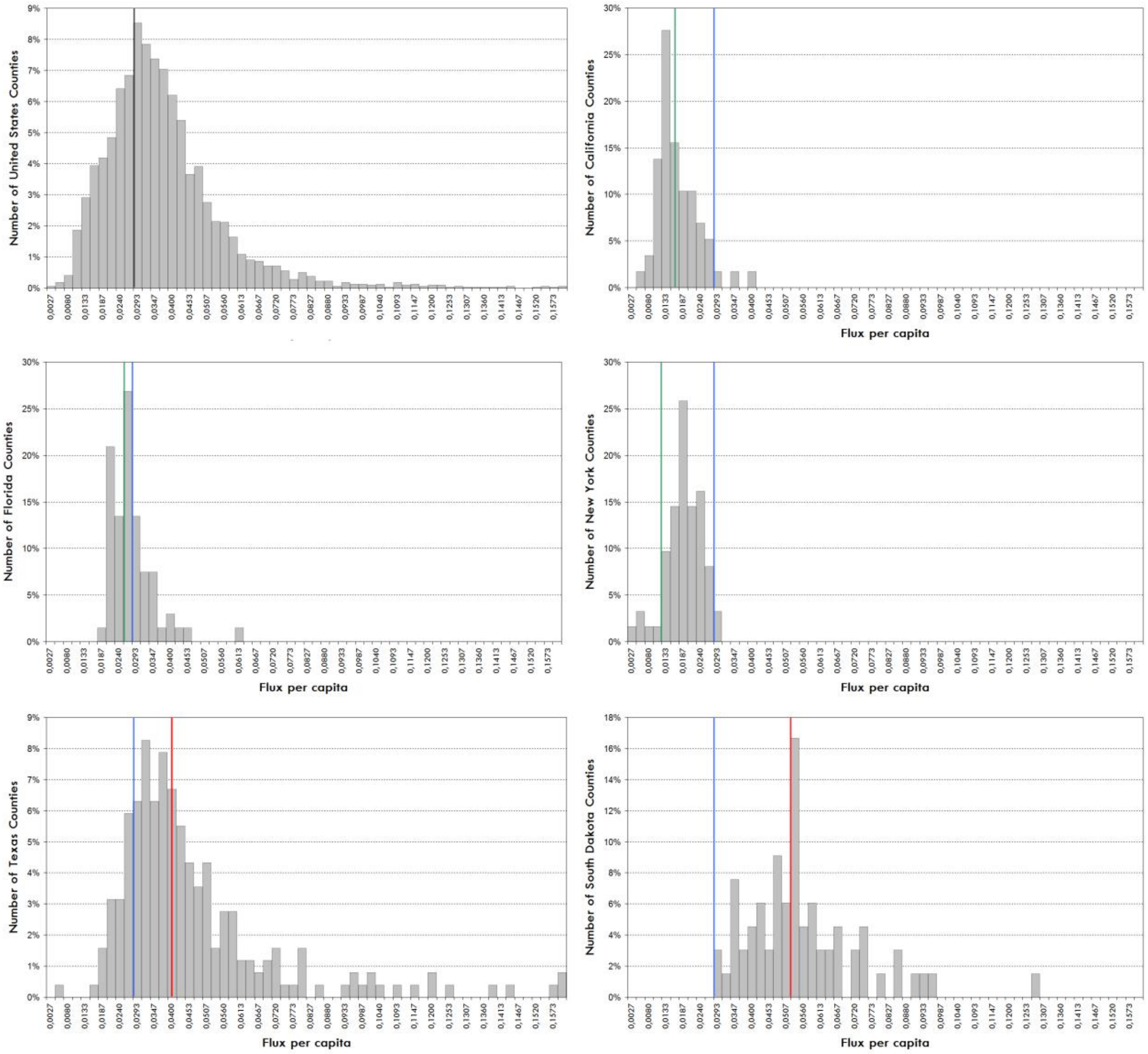

**Figure 9. Flux per capita (in arbitrary units, $\propto$ W $sr^{-1}$ per capita, see main text) in USA (upper left graph), and in 5 US States. The blue line indicates the average USA flux per capita. The green (if better) or red (if worse) lines indicate the State's average. . The histograms show the percent of counties in USA and California, Florida, New York, Texas and South Dakota counties in each flux per capita range. From left to right the pollution produced per capita increases.**

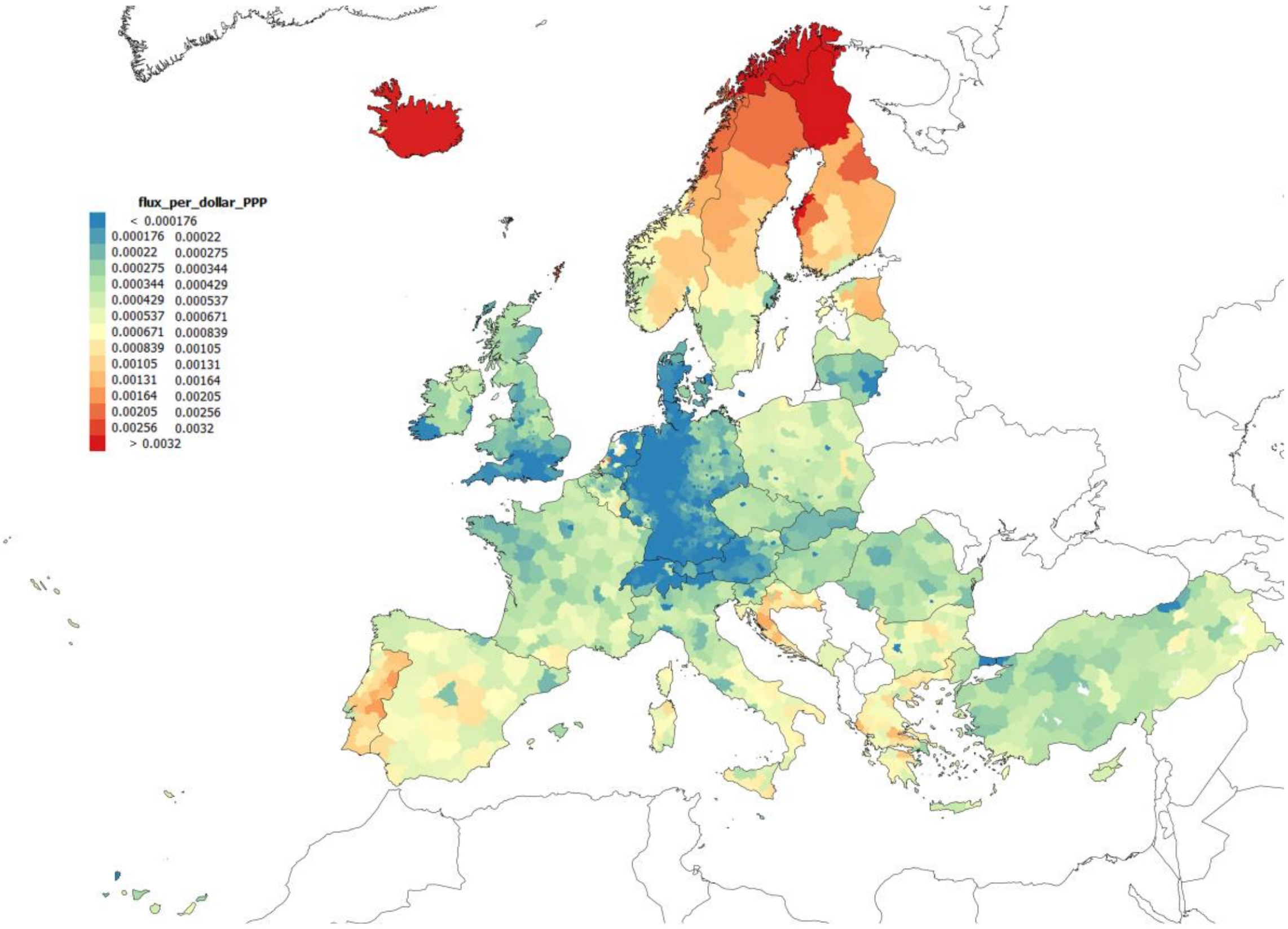

**Figure 10. Flux per GDP unit in Europe's NUTS3 (arbitrary units, $\propto$ W sr$^{-1}$ \$$^{-1}$ see main text). This map shows how diffferent can be the level of light produced per each units of Gross Domestic Product, so that light pollution is not directly binned to the wealth of a country's economy (e.g., compare Germany and Switzerland to Portugal and Greece).**

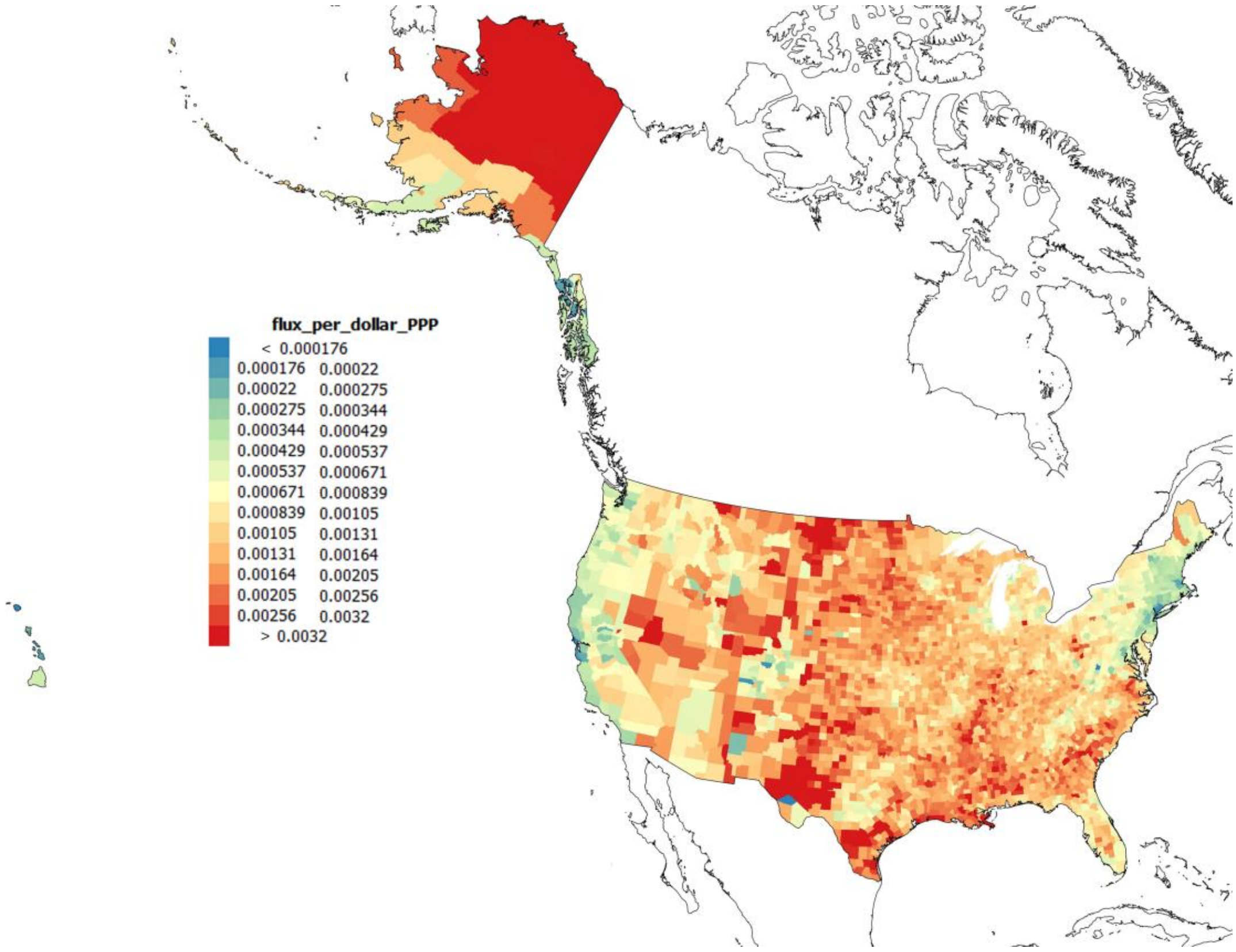


**Figure 11. Flux per GDP in US counties (arbitrary units, $\propto$ W sr$^{-1}$ \$$^{-1}$, see main text).**

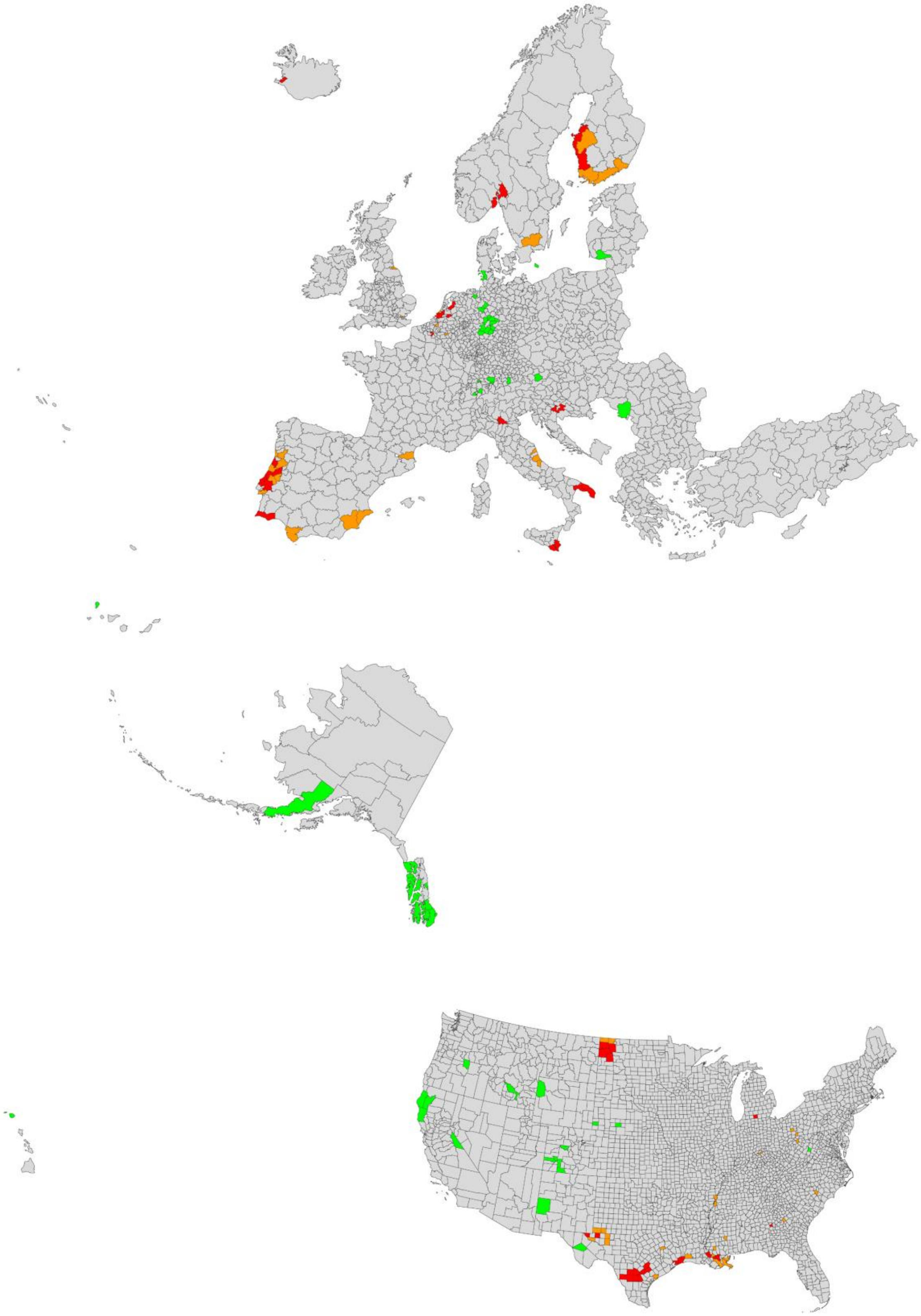

**Figure 12. In green the 'good' NUTS3 administrative regions (upper map) and counties (lower map), in orange the 'bad', second to last in the global ranking (26th to 50th last places) and the red 'ugly' (very last to 25th place from bottom).**

| NUTS_ID | NUTS3 Name | Country | Flux per Capita ( ∝W $sr^{-1}$) | Rank FpC | GDP PPP per capita 2014 | Flux per dollar PPP ( ∝ W $sr^{-1}$ $\$^{-1}$) | rank FpD | Mean radiance (mcd/$m^2$) | rank radiance | Sum of Ranks | Global rank |
|---|---|---|---|---|---|---|---|---|---|---|---|
| DE946 | Ammerland | Germany | 2,89 | 7 | 31.489 | 0,0918 | 78 | 0,103 | 294 | 379 | 1 |
| DK014 | Bornholm | Denmark | 4,79 | 167 | 32.157 | 0,149 | 240 | 0,0310 | 21 | 428 | 2 |
| DE918 | Northeim | Germany | 4,03 | 68 | 30.153 | 0,134 | 194 | 0,0789 | 176 | 438 | 3 |
| DE724 | Marburg-Biedenkopf | Germany | 3,44 | 27 | 39.512 | 0,0870 | 67 | 0,115 | 355 | 449 | 4 |
| CH063 | Schwyz | Switzerland | 2,62 | 3 | 40230 | 0,0650 | 30 | 0,129 | 419 | 452 | 5 |
| ES707 | La Palma | Spain | 4,12 | 80 | 23.106 | 0,178 | 337 | 0,0431 | 46 | 463 | 6 |
| DE915 | Göttingen | Germany | 3,33 | 18 | 39.165 | 0,0850 | 60 | 0,121 | 385 | 463 | 7 |
| DE919 | Osterode am Harz | Germany | 4,28 | 104 | 34.457 | 0,124 | 166 | 0,0828 | 196 | 466 | 8 |
| DE216 | Bad Tölz-Wolfratshausen | Germany | 3,92 | 57 | 32.382 | 0,121 | 152 | 0,0962 | 258 | 467 | 9 |
| DE926 | Holzminden | Germany | 4,22 | 94 | 31.581 | 0,134 | 195 | 0,0806 | 183 | 472 | 10 |
| DE148 | Ravensburg | Germany | 4,09 | 77 | 43.856 | 0,0932 | 80 | 0,109 | 326 | 483 | 11 |
| DE735 | Schwalm-Eder-Kreis | Germany | 4,23 | 95 | 33.256 | 0,127 | 173 | 0,0909 | 234 | 502 | 12 |
| DEF07 | Nordfriesland | Germany | 5,24 | 233 | 38.120 | 0,137 | 206 | 0,0523 | 67 | 506 | 13 |
| LT007 | Tauragės apskritis | Lithuania | 3,45 | 28 | 14.088 | 0,245 | 499 | 0,0191 | 5 | 532 | 14 |
| DE27A | Lindau (Bodensee) | Germany | 3,69 | 35 | 40.267 | 0,0918 | 76 | 0,133 | 438 | 549 | 15 |
| DE737 | Werra-Meißner-Kreis | Germany | 4,37 | 110 | 27.922 | 0,157 | 272 | 0,0789 | 175 | 557 | 16 |
| DEA46 | Minden-Lübbecke | Germany | 3,39 | 22 | 44.299 | 0,0766 | 45 | 0,157 | 517 | 584 | 17 |
| DE736 | Waldeck-Frankenberg | Germany | 5,19 | 228 | 37.834 | 0,137 | 205 | 0,0742 | 156 | 589 | 18 |
| CH064 | Obwalden | Switzerland | 5,35 | 250 | 44543 | 0,120 | 149 | 0,0821 | 191 | 590 | 19 |
| CH052 | Schaffhausen | Switzerland | 2,22 | 1 | 59470 | 0,0374 | 3 | 0,176 | 589 | 593 | 20 |
| AT314 | Steyr-Kirchdorf | Austria | 5,77 | 304 | 46.573 | 0,124 | 162 | 0,0678 | 128 | 594 | 21 |
| DEA44 | Höxter | Germany | 4,43 | 115 | 29.690 | 0,149 | 243 | 0,0926 | 241 | 599 | 22 |
| RO422 | Caraş-Severin | Romania | 3,71 | 36 | 13.519 | 0,275 | 555 | 0,0316 | 22 | 613 | 23 |
| DE733 | Hersfeld-Rotenburg | Germany | 5,26 | 236 | 40.898 | 0,129 | 175 | 0,0869 | 216 | 627 | 24 |
| DE927 | Nienburg (Weser) | Germany | 4,86 | 178 | 34.360 | 0,142 | 224 | 0,0887 | 228 | 630 | 25 |
| | | | | | | | | | | | |
| ITF13 | Pescara | Italy | 19,1 | 1188 | 29.739 | 0,642 | 1078 | 0,716 | 1110 | 3376 | 1310 |
| FI194 | Etelä-Pohjanmaa | Finland | 73,4 | 1348 | 30.787 | 2,38 | 1349 | 0,215 | 691 | 3388 | 1311 |
| PT165 | Dão-Lafões | Portugal | 22,4 | 1242 | 20.005 | 1,12 | 1286 | 0,331 | 863 | 3391 | 1312 |
| BE334 | Arr. Waremme | Belgium | 16,3 | 1093 | 19.120 | 0,851 | 1206 | 0,685 | 1093 | 3392 | 1313 |
| ES612 | Cádiz | Spain | 19,8 | 1200 | 21.348 | 0,926 | 1237 | 0,464 | 968 | 3405 | 1314 |
| FR815 | Pyrénées-Orientales | France | 23,5 | 1258 | 24.452 | 0,961 | 1245 | 0,390 | 907 | 3410 | 1315 |
| FI1C2 | Kanta-Häme | Finland | 43,4 | 1328 | 31.623 | 1,37 | 1318 | 0,253 | 768 | 3414 | 1316 |
| PT115 | Tâmega | Portugal | 15,7 | 1077 | 16.352 | 0,961 | 1246 | 0,700 | 1100 | 3423 | 1317 |
| PT166 | Pinhal Interior Sul | Portugal | 38,2 | 1323 | 21.773 | 1,76 | 1342 | 0,255 | 772 | 3437 | 1318 |
| BE236 | Arr. Sint-Niklaas | Belgium | 19,1 | 1186 | 35.457 | 0,538 | 973 | 1,91 | 1285 | 3444 | 1319 |
| UKC11 | Hartlepool and Stockton-on-Tees | United Kingdom | 16,7 | 1111 | 28.173 | 0,592 | 1025 | 2,18 | 1308 | 3444 | 1320 |
| UKH32 | Thurrock | United Kingdom | 16,7 | 1113 | 28.072 | 0,595 | 1029 | 2,29 | 1313 | 3455 | 1321 |
| FI1C4 | Kymenlaakso | Finland | 52,2 | 1340 | 32.712 | 1,60 | 1331 | 0,268 | 793 | 3464 | 1322 |
| ES620 | Murcia | Spain | 23,0 | 1253 | 24.986 | 0,921 | 1235 | 0,474 | 976 | 3464 | 1323 |
| NL332 | Agglomeratie 's-Gravenhage | Nederlands | 21,7 | 1231 | 46.972 | 0,462 | 875 | 13,6 | 1358 | 3464 | 1324 |
| PT16C | Médio Tejo | Portugal | 27,6 | 1286 | 21.773 | 1,27 | 1303 | 0,362 | 883 | 3472 | 1325 |
| PT114 | Grande Porto | Portugal | 16,1 | 1086 | 24.557 | 0,655 | 1097 | 1,96 | 1290 | 3473 | 1326 |
| UKC12 | South Teesside | United Kingdom | 16,4 | 1100 | 24.505 | 0,671 | 1108 | 1,65 | 1266 | 3474 | 1327 |
| ITI35 | Fermo | Italy | 22,0 | 1236 | 29.291 | 0,752 | 1164 | 0,654 | 1075 | 3475 | 1328 |
| FI1C1 | Varsinais-Suomi | Finland | 45,5 | 1330 | 34.670 | 1,31 | 1312 | 0,308 | 835 | 3477 | 1329 |
| PT162 | Baixo Mondego | Portugal | 23,9 | 1265 | 23.197 | 1,03 | 1270 | 0,432 | 943 | 3478 | 1330 |
| ITF12 | Teramo | Italy | 22,8 | 1250 | 27.776 | 0,821 | 1188 | 0,568 | 1045 | 3483 | 1331 |
| PT172 | Península de Setúbal | Portugal | 19,3 | 1193 | 35.315 | 0,546 | 981 | 2,22 | 1311 | 3485 | 1332 |
| FI1B1 | Helsinki-Uusimaa | Finland | 31,1 | 1307 | 49.670 | 0,627 | 1057 | 0,760 | 1123 | 3487 | 1333 |
| ES521 | Alicante / Alacant | Spain | 18,3 | 1163 | 24.095 | 0,758 | 1168 | 0,872 | 1157 | 3488 | 1334 |
| ITC4B | Mantova | Italy | 23,3 | 1254 | 35.729 | 0,652 | 1093 | 0,813 | 1144 | 3491 | 1335 |
| FI196 | Satakunta | Finland | 65,0 | 1347 | 35.485 | 1,83 | 1345 | 0,276 | 800 | 3492 | 1336 |
| ITF44 | Brindisi | Italy | 19,1 | 1191 | 21.253 | 0,900 | 1225 | 0,659 | 1077 | 3493 | 1337 |
| PT163 | Pinhal Litoral | Portugal | 24,0 | 1266 | 26.153 | 0,918 | 1233 | 0,507 | 1000 | 3499 | 1338 |
| BE321 | Arr. Ath | Belgium | 19,0 | 1183 | 22.206 | 0,856 | 1211 | 0,733 | 1114 | 3508 | 1339 |
| ITG19 | Siracusa | Italy | 20,4 | 1209 | 18.363 | 1,11 | 1283 | 0,547 | 1027 | 3519 | 1340 |
| ES640 | Melilla | Spain | 15,6 | 1066 | 22.879 | 0,680 | 1115 | 5,02 | 1348 | 3529 | 1341 |
| IS001 | Höfuðborgarsvæði (Capital Region) | Iceland | 29,5 | 1299 | 40953 | 0,720 | 1145 | 0,674 | 1086 | 3530 | 1342 |
| PT185 | Lezíria do Tejo | Portugal | 31,0 | 1306 | 21.812 | 1,42 | 1320 | 0,392 | 910 | 3536 | 1343 |
| NO033 | Vestfold | Norway | 32,7 | 1310 | 34.126 | 0,958 | 1244 | 0,483 | 983 | 3537 | 1344 |
| PT164 | Pinhal Interior Norte | Portugal | 27,2 | 1285 | 23.197 | 1,17 | 1295 | 0,450 | 958 | 3538 | 1345 |
| ITF45 | Lecce | Italy | 18,3 | 1164 | 18.916 | 0,968 | 1248 | 0,812 | 1142 | 3554 | 1346 |

| | | | | | | | | | | | | |
|---|---|---|---|---|---|---|---|---|---|---|---|---|
| PT150 | Algarve | Portugal | 34,8 | 1315 | 26.639 | 1,31 | 1308 | 0,422 | 934 | 3557 | 1347 |
| PT161 | Baixo Vouga | Portugal | 21,2 | 1224 | 24.821 | 0,852 | 1208 | 0,764 | 1126 | 3558 | 1348 |
| ITG18 | Ragusa | Italy | 22,4 | 1245 | 21.871 | 1,03 | 1267 | 0,581 | 1049 | 3561 | 1349 |
| HR042 | Zagrebačka županija | Croatia | 23,5 | 1259 | 15.059 | 1,56 | 1330 | 0,520 | 1009 | 3598 | 1350 |
| ITF43 | Taranto | Italy | 20,9 | 1221 | 21.020 | 0,995 | 1261 | 0,741 | 1118 | 3600 | 1351 |
| NO012 | Akershus | Norway | 37,4 | 1322 | 43.755 | 0,855 | 1210 | 0,676 | 1088 | 3620 | 1352 |
| NL224 | Zuidwest-Gelderland | Nederlands | 26,8 | 1281 | 37.555 | 0,713 | 1139 | 1,45 | 1247 | 3667 | 1353 |
| PT16B | Oeste | Portugal | 28,5 | 1292 | 21.457 | 1,33 | 1315 | 0,781 | 1137 | 3744 | 1354 |
| NL230 | Flevoland | Nederlands | 34,8 | 1316 | 33.843 | 1,03 | 1272 | 0,969 | 1183 | 3771 | 1355 |
| NL339 | Groot-Rijnmond | Nederlands | 32,8 | 1311 | 46.753 | 0,702 | 1127 | 6,26 | 1351 | 3789 | 1356 |
| FI195 | Pohjanmaa | Finland | 267 | 1359 | 38.280 | 6,98 | 1359 | 0,796 | 1139 | 3857 | 1357 |
| NL338 | Oost-Zuid-Holland | Nederlands | 88,5 | 1350 | 37.039 | 2,39 | 1350 | 7,15 | 1355 | 4055 | 1358 |
| NL333 | Delft en Westland | Nederlands | 199 | 1356 | 48.328 | 4,11 | 1356 | 30,0 | 1359 | 4071 | 1359 |

Table 1. The Global rankings Top 25 (in green, the good), and bottom 50 to 26 (in orange, the bad) and bottom 25 (in red, the ugly) NUTS3 administrative units in Europe. The values in the Flux per Capita column was multiplied by $10^3$ and the values in the Flux per dollar by $10^6$ in order to get more readable numbers.

| GEOID1 | Name, State | Flux per Capita (∝W $sr^{-1}$) | rank FpC | GDP PPP per capita 2014 | Flux per dollar PPP (∝ W $sr^{-1}$ $\$^{-1}$) | rank fpd | Mean radiance ($mcd/m^2$) | rank radiance | Sum of Ranks | Global rank |
|---|---|---|---|---|---|---|---|---|---|---|
| 0500000US02105 | Hoonah-Angoon Census Area, Alaska | 7,24 | 14 | 30811 | 0,235 | 19 | 0,000429 | 8 | 41 | 1 |
| 0500000US35003 | Catron County, New Mexico | 5,76 | 9 | 19254 | 0,299 | 38 | 0,000678 | 12 | 59 | 2 |
| 0500000US48243 | Jeff Davis County, Texas | 4,95 | 7 | 28902 | 0,171 | 8 | 0,00256 | 62 | 77 | 3 |
| 0500000US08113 | San Miguel County, Colorado | 8,49 | 22 | 40993 | 0,207 | 14 | 0,00199 | 47 | 83 | 4 |
| 0500000US15005 | Kalawao County, Hawaii | 0 | 1 | 43771 | 0 | 1 | 0,00403 | 102 | 104 | 5 |
| 0500000US02275 | Wrangell City and Borough, Alaska | 11,1 | 52 | 30671 | 0,361 | 68 | 0,000410 | 7 | 127 | 6 |
| 0500000US06105 | Trinity County, California | 7,98 | 16 | 23145 | 0,345 | 58 | 0,00326 | 84 | 158 | 7 |
| 0500000US02198 | Prince of Wales-Hyder Census Area, Alaska | 10,6 | 43 | 24737 | 0,429 | 114 | 0,00119 | 24 | 181 | 8 |
| 0500000US08053 | Hinsdale County, Colorado | 13,6 | 124 | 36046 | 0,376 | 75 | 0,00131 | 27 | 226 | 9 |
| 0500000US31117 | McPherson County, Nebraska | 10,7 | 46 | 25760 | 0,416 | 100 | 0,00331 | 85 | 231 | 10 |
| 0500000US51091 | Highland County, Virginia | 4,14 | 3 | 26949 | 0,154 | 6 | 0,00994 | 280 | 289 | 11 |
| 0500000US16013 | Blaine County, Idaho | 12,6 | 89 | 34517 | 0,366 | 71 | 0,00518 | 136 | 296 | 12 |
| 0500000US41069 | Wheeler County, Oregon | 13,0 | 100 | 24154 | 0,539 | 186 | 0,000907 | 17 | 303 | 13 |
| 0500000US06051 | Mono County, California | 13,0 | 102 | 29578 | 0,441 | 120 | 0,00313 | 81 | 303 | 14 |
| 0500000US02130 | Ketchikan Gateway Borough, Alaska | 14,4 | 154 | 31494 | 0,456 | 125 | 0,00192 | 44 | 323 | 15 |
| 0500000US08091 | Ouray County, Colorado | 13,3 | 114 | 32562 | 0,409 | 96 | 0,00454 | 121 | 331 | 16 |
| 0500000US02164 | Lake and Peninsula Borough, Alaska | 12,7 | 91 | 21581 | 0,587 | 235 | 0,000683 | 13 | 339 | 17 |
| 0500000US56039 | Teton County, Wyoming | 15,5 | 205 | 43628 | 0,356 | 63 | 0,00373 | 93 | 361 | 18 |
| 0500000US08097 | Pitkin County, Colorado | 11,1 | 53 | 54441 | 0,205 | 13 | 0,0112 | 305 | 371 | 19 |
| 0500000US31007 | Banner County, Nebraska | 10,2 | 38 | 33226 | 0,306 | 41 | 0,0121 | 328 | 407 | 20 |
| 0500000US08007 | Archuleta County, Colorado | 13,8 | 139 | 28506 | 0,486 | 144 | 0,00561 | 146 | 429 | 21 |
| 0500000US06045 | Mendocino County, California | 8,70 | 25 | 23712 | 0,367 | 73 | 0,0127 | 342 | 440 | 22 |
| 0500000US02220 | Sitka City and Borough, Alaska | 17,2 | 281 | 33920 | 0,507 | 167 | 0,000610 | 11 | 459 | 23 |
| 0500000US06023 | Humboldt County, California | 8,01 | 17 | 23516 | 0,341 | 57 | 0,0159 | 402 | 476 | 24 |
| 0500000US15007 | Kauai County, Hawaii | 4,98 | 8 | 27079 | 0,184 | 10 | 0,0205 | 477 | 495 | 25 |
| | | | | | | | | | | |
| 0500000US22091 | St. Helena Parish, Louisiana | 94,4 | 3044 | 19582 | 4,82 | 3066 | 0,206 | 2173 | 8283 | 3093 |
| 0500000US48173 | Glasscock County, Texas | 1387 | 3133 | 39169 | 35,4 | 3129 | 0,174 | 2027 | 8289 | 3094 |
| 0500000US28143 | Tunica County, Mississippi | 91,7 | 3039 | 15298 | 6,00 | 3090 | 0,204 | 2161 | 8290 | 3095 |
| 0500000US48313 | Madison County, Texas | 108 | 3061 | 15222 | 7,08 | 3099 | 0,196 | 2138 | 8298 | 3096 |
| 0500000US13289 | Twiggs County, Georgia | 76,8 | 2990 | 17629 | 4,36 | 3051 | 0,242 | 2311 | 8352 | 3097 |
| 0500000US48383 | Reagan County, Texas | 1010 | 3131 | 23814 | 42,4 | 3131 | 0,185 | 2093 | 8355 | 3098 |
| 0500000US28031 | Covington County, Mississippi | 83,0 | 3021 | 16733 | 4,96 | 3072 | 0,230 | 2270 | 8363 | 3099 |
| 0500000US21077 | Gallatin County, Kentucky | 66,6 | 2917 | 20507 | 3,25 | 2971 | 0,311 | 2477 | 8365 | 3100 |
| 0500000US48475 | Ward County, Texas | 304 | 3116 | 23887 | 12,7 | 3119 | 0,193 | 2131 | 8366 | 3101 |
| 0500000US22057 | Lafourche Parish, Louisiana | 68,5 | 2936 | 25010 | 2,74 | 2851 | 0,386 | 2594 | 8381 | 3102 |
| 0500000US54017 | Doddridge County, West Virginia | 172 | 3104 | 18552 | 9,26 | 3110 | 0,206 | 2174 | 8388 | 3103 |
| 0500000US45069 | Marlboro County, South Carolina | 67,5 | 2929 | 14925 | 4,52 | 3056 | 0,287 | 2431 | 8416 | 3104 |
| 0500000US22087 | St. Bernard Parish, Louisiana | 73,3 | 2978 | 21079 | 3,48 | 2997 | 0,295 | 2448 | 8423 | 3105 |
| 0500000US38023 | Divide County, North Dakota | 2065 | 3135 | 41003 | 50,4 | 3134 | 0,211 | 2189 | 8458 | 3106 |
| 0500000US48003 | Andrews County, Texas | 337 | 3119 | 29363 | 11,5 | 3114 | 0,225 | 2248 | 8481 | 3107 |
| 0500000US54051 | Marshall County, West Virginia | 72,0 | 2966 | 24419 | 2,95 | 2919 | 0,389 | 2604 | 8489 | 3108 |
| 0500000US22075 | Plaquemines Parish, Louisiana | 150 | 3090 | 26672 | 5,62 | 3085 | 0,248 | 2327 | 8502 | 3109 |
| 0500000US48057 | Calhoun County, Texas | 137 | 3086 | 24142 | 5,68 | 3086 | 0,250 | 2335 | 8507 | 3110 |
| 0500000US21041 | Carroll County, Kentucky | 75,6 | 2984 | 19711 | 3,83 | 3027 | 0,334 | 2518 | 8529 | 3111 |
| 0500000US38013 | Burke County, North Dakota | 1465 | 3134 | 33174 | 44,2 | 3132 | 0,236 | 2294 | 8560 | 3112 |
| 0500000US22019 | Calcasieu Parish, Louisiana | 66,3 | 2910 | 24521 | 2,70 | 2838 | 0,803 | 2899 | 8647 | 3113 |
| 0500000US48317 | Martin County, Texas | 943 | 3129 | 26286 | 35,9 | 3130 | 0,276 | 2404 | 8663 | 3114 |
| 0500000US39067 | Harrison County, Ohio | 153 | 3091 | 22180 | 6,90 | 3098 | 0,324 | 2501 | 8690 | 3115 |
| 0500000US22005 | Ascension Parish, Louisiana | 71,5 | 2960 | 28834 | 2,48 | 2750 | 1,56 | 3030 | 8740 | 3116 |
| 0500000US05035 | Crittenden County, Arkansas | 74,5 | 2983 | 19732 | 3,78 | 3023 | 0,529 | 2740 | 8746 | 3117 |
| 0500000US48163 | Frio County, Texas | 252 | 3112 | 15732 | 16,0 | 3121 | 0,331 | 2515 | 8748 | 3118 |
| 0500000US48297 | Live Oak County, Texas | 433 | 3122 | 21357 | 20,3 | 3124 | 0,325 | 2503 | 8749 | 3119 |
| 0500000US13053 | Chattahoochee County, Georgia | 122 | 3077 | 19538 | 6,25 | 3092 | 0,401 | 2621 | 8790 | 3120 |
| 0500000US48177 | Gonzales County, Texas | 251 | 3111 | 20794 | 12,1 | 3117 | 0,366 | 2564 | 8792 | 3121 |
| 0500000US48361 | Orange County, Texas | 70,9 | 2953 | 24938 | 2,84 | 2886 | 1,20 | 2982 | 8821 | 3122 |
| 0500000US48135 | Ector County, Texas | 84,0 | 3023 | 25726 | 3,27 | 2976 | 0,642 | 2828 | 8827 | 3123 |
| 0500000US22095 | St. John the Baptist Parish, Louisiana | 77,2 | 2991 | 22785 | 3,39 | 2990 | 0,709 | 2862 | 8843 | 3124 |
| 0500000US48301 | Loving County, Texas | 41668 | 3142 | 25629 | 1626 | 3142 | 0,399 | 2618 | 8902 | 3125 |
| 0500000US22047 | Iberville Parish, Louisiana | 108 | 3063 | 21576 | 5,02 | 3073 | 0,608 | 2810 | 8946 | 3126 |
| 0500000US48245 | Jefferson County, Texas | 76,8 | 2989 | 23563 | 3,26 | 2974 | 1,25 | 2988 | 8951 | 3127 |
| 0500000US48013 | Atascosa County, Texas | 186 | 3107 | 21957 | 8,46 | 3107 | 0,525 | 2738 | 8952 | 3128 |
| 0500000US38025 | Dunn County, North Dakota | 3720 | 3139 | 38216 | 97,4 | 3138 | 0,458 | 2687 | 8964 | 3129 |
| 0500000US22093 | St. James Parish, Louisiana | 108 | 3062 | 24757 | 4,36 | 3052 | 0,701 | 2856 | 8970 | 3130 |
| 0500000US48071 | Chambers County, Texas | 156 | 3093 | 30102 | 5,19 | 3077 | 0,611 | 2814 | 8984 | 3131 |
| 0500000US48123 | DeWitt County, Texas | 445 | 3123 | 26104 | 17,1 | 3122 | 0,571 | 2789 | 9034 | 3132 |
| 0500000US26023 | Branch County, Michigan | 135 | 3085 | 20823 | 6,50 | 3094 | 0,701 | 2857 | 9036 | 3133 |
| 0500000US48127 | Dimmit County, Texas | 1208 | 3132 | 20822 | 58,0 | 3136 | 0,582 | 2796 | 9064 | 3134 |
| 0500000US22089 | St. Charles Parish, Louisiana | 107 | 3058 | 26623 | 4,00 | 3035 | 1,25 | 2985 | 9078 | 3135 |
| 0500000US38105 | Williams County, North Dakota | 967 | 3130 | 41984 | 23,0 | 3126 | 0,675 | 2838 | 9094 | 3136 |
| 0500000US38061 | Mountrail County, North Dakota | 2199 | 3136 | 33839 | 65,0 | 3137 | 0,672 | 2837 | 9110 | 3137 |
| 0500000US22121 | West Baton Rouge Parish, Louisiana | 140 | 3087 | 25296 | 5,52 | 3083 | 1,40 | 3015 | 9185 | 3138 |
| 0500000US48283 | La Salle County, Texas | 2360 | 3137 | 17184 | 137 | 3139 | 0,848 | 2913 | 9189 | 3139 |
| 0500000US48255 | Karnes County, Texas | 790 | 3128 | 22966 | 34,4 | 3128 | 0,984 | 2938 | 9194 | 3140 |
| 0500000US48311 | McMullen County, Texas | 13706 | 3141 | 36277 | 378 | 3141 | 0,899 | 2919 | 9201 | 3141 |
| 0500000US38053 | McKenzie County, North Dakota | 9138 | 3140 | 34688 | 263 | 3140 | 1,32 | 2999 | 9279 | 3142 |

Table 2. The Global rankings Top 25 (in green, the good), and bottom 50 to 26 (in orange, the bad) and bottom 25 (in red, the ugly) US Counties. The values in the Flux per Capita column was multiplied by $10^3$ and the values in the Flux per dollar by $10^6$ in order to get more readable numbers.

**Acknowledgments**

**Funding:** No special funds were used for the research that carried to this publication

**Author contributions:** F.F. and R.F. conceived the research, F.F. wrote most of the manuscript, N.R. and B.P. performed the statistics of population and territory using the World Atlas sky brightness data, F.F. and R.F. performed the QGIS analysis and produced the maps, graphs and tables, T.G. performed the economic data search and analysis and wrote the pertaining parts. K.B. and C.D.E. provided the radiance data. All authors read and approved the manuscript.

**Competing interests:** The authors declare to have no competing interests. Notwithstanding this, F.F. retains correct to say that he's president of CieloBuio, an Italian association for the protection of the night sky.

[18] Gallaway, Terrel. 2010. "On Light Pollution, Passive Pleasures, and the Instrumental Value of Beauty." Journal of Economic Issues 44 (1):71-88. doi: https://doi.org/10.2753/JEI0021-3624440104
[19] Meier, Josiane, Ute Hasenöhrl, Katharina Krause, and Merle Pottharst. 2015. Urban lighting, light pollution, and society. New York: Routledge, Taylor & Francis Group.
[20] Aubrecht, Christoph, Chris Elvidge, Daniel Ziskin, Pedro Rodrigues, and Artur Gil. 2010. "Observing stress of artificial night lighting on marine ecosystems-a remote sensing application study." International Archives of the Photogrammetry, Remote Sensing and Spatial Information Science Technical Commission VII Symposium, 100 Years ISPRS, Advancing Remote Sensing Science, Vienna, Austria.
[21] Pengpeng, Han, Huang Jinliang, Li Rendong, Wang Lihui, Hu Yanxia, Wang Jiuling, and Huang Wei. 2014. "Monitoring Trends in Light Pollution in China Based on Nighttime Satellite Imagery." Remote Sensing 6 (6):5541-5558. doi: 10.3390/rs6065541.
[22] Da Silva, Arnaud, Jelmer M. Samplonius, Emmi Schlicht, Mihai Valcu, and Bart Kempenaers. 2014. "Artificial night lighting rather than traffic noise affects the daily timing of dawn and dusk singing in common European songbirds." Behavioral Ecology 25 (5):1037-1047. doi: 10.1093/beheco/aru103.
[23] Kwak, Myeong Ja, Sun Mi Je, Hyo Cheng Cheng, Se Myeong Seo, Jeong Ho Park, Saeng Geul Baek, Inkyin Khaine, Taeyoon Lee, Jihwi Jang, Yang Li, Haenaem Kim, Jong Kyu Lee, Jieun Kim, and Su Young Woo. 2018. "Night Light-Adaptation Strategies for Photosynthetic Apparatus in Yellow-Poplar (Liriodendron tulipifera L.) Exposed to Artificial Night Lighting." Forests (19994907) 9 (2):74. doi: 10.3390/f9020074.
[24] Gallaway, Terrel A., Reed Neil Olsen, and David M. Mitchell. 2013. "Blinded by the Light: Economic Analysis of Severe Light Pollution." Journal of Economics (MVEA) 39 (1):45-63. doi: http://www.cba.uni.edu/economics/joe.htm.
[25] Gallaway, Terrel, Reed N. Olsen, and David M. Mitchell. 2010. "The Economics of Global Light Pollution." Ecological Economics 69 (3):658-665. doi: 10.1016/j.ecolecon.2009.10.003.
[26] Eisenbeis, Gerhard, Andreas Hänel, M McDonnell, A Hahs, and J Breuste. 2009. "Light pollution and the impact of artificial night lighting on insects." Ecology of cities and towns:243-263. doi: 10.1017/CBO9780511609763.016.
[27] Geffen, Koert G., Emiel Eck, Rens A. Boer, Roy H. A. Grunsven, Lucia Salis, Frank Berendse, Elmar M. Veenendaal, Alan Stewart, and Steven Sait. 2015. "Artificial light at night inhibits mating in a Geometrid moth." Insect Conservation & Diversity 8 (3):282-287. doi: 10.1111/icad.12116.
[28] Van Langevelde, Frank, Roy HA Van Grunsven, Elmar M Veenendaal, and Thijs PM Fijen. 2017. "Artificial night lighting inhibits feeding in moths." Biology Letters 13 (3):20160874. doi: https://doi.org/10.1098/rsbl.2016.0874.
[29] Kempenaers, Bart, Pernilla Borgstroem, Peter Loes, Emmi Schlicht, and Mihai Valcu. 2010. "Artificial Night Lighting Affects Dawn Song, Extra-Pair Siring Success, and Lay Date in Songbirds." Current Biology 20 (19):1735-1739. doi: 10.1016/j.cub.2010.08.028.
[30] Mathews, Fiona, Niamh Roche, Tina Aughney, Nicholas Jones, Julie Day, James Baker, and Steve Langton. 2015. "Barriers and benefits: implications of artificial night-lighting for the distribution of common bats in Britain and Ireland." Philosophical Transactions of the Royal Society B: Biological Sciences 370 (1667):1-13. doi: 10.1098/rstb.2014.0124.
[31] Rotics, Shay, Tamar Dayan, and Noga Kronfeld-Schor. 2011. "Effect of artificial night lighting on temporally partitioned spiny mice." Journal of Mammalogy 92 (1):159-168. doi: 10.1644/10-MAMM-A-112.1.
[32] Salmon, Michael. 2003. "Artificial night lighting and sea turtles." Biologist 50 (4):163-168.

[33] Elvidge, C.D.; Ziskin, D.; Baugh, K.E.; Tuttle, B.T.; Ghosh, T.; Pack, D.W. A fifteen year record of global natural gas flaring derived from satellite data. Energies 2009, 2, 595–622.

[34] Bouroussis, Constantinos A, and Frangiskos V Topalis. 2016. "Smart multi-workplane lighting control and utilization of daylight using an imaging photosensor." 2016 IEEE 16th International Conference on Environment and Electrical Engineering (EEEIC). doi: 10.1109/EEEIC.2016.7555530

[35] Kontadakis, Antonis, Aris Tsangrassoulis, L. Doulos, and F. Topalis. 2017. "An active sunlight redirection system for daylight enhancement beyond the perimeter zone."  Building & Environment 113:267-279. doi: 10.1016/j.buildenv.2016.09.029

[36] EUROSTAT, https://ec.europa.eu/eurostat/web/gisco/geodata/reference-data/administrative-units-statistical-units/nuts

[37] United States Census Bureau, https://www.census.gov/geo/maps-data/data/cbf/cbf_counties.html

[38] F. Falchi, P. Cinzano, D. Duriscoe, C. C. M. Kyba, C. D. Elvidge, K. Baugh, B. Portnov, N. A. Rybnikova, R. Furgoni, Supplement to the New World Atlas of Artificial Night Sky Brightness, GFZ Data Services (2016); http://doi.org/10.5880/GFZ.1.4.2016.001

[39] Oak Ridge National Laboratory, LandScan™, http://web.ornl.gov/sci/landscan/ [accessed November 11, 2018]

[40] Organization for Economic Cooperation and Development, http://www.oecd.org/cfe/regional-policy/regionalstatisticsandindicators.htm

[41] United States Census Bureau, https://factfinder.census.gov/